\documentclass[journal, 12pt, draftcls, onecolumn ]{IEEEtran}
\usepackage[top=1.2in, bottom=1.2in, left=1in, right=1in]{geometry}
\usepackage{cite}
\usepackage{footnote}
\usepackage{amsmath,amsfonts,amssymb,graphicx}
\usepackage{graphicx}
\usepackage{ifpdf}
\usepackage{amsmath}
\ifCLASSINFOpdf
\else
\fi

\begin{document}
\setlength\arraycolsep{1.5pt}
\bibliographystyle{IEEEtran}  
\title{Optimization of UAV Heading \\ for the Ground-to-Air Uplink}

\author{\IEEEauthorblockN{Feng Jiang and A. Lee Swindlehurst}\\
\IEEEauthorblockA{Department of Electrical Engineering and Computer Science\\University of California, Irvine\\Irvine, CA, 92697, USA\\Email:\{feng.jiang, swindle\}@uci.edu\\}}


%


\maketitle

\begin{abstract}
In this paper we consider a collection of single-antenna ground nodes
communicating with a multi-antenna unmanned aerial vehicle (UAV) over
a multiple-access ground-to-air wireless communications link.  The UAV
uses beamforming to mitigate the inter-user interference and achieve
spatial division multiple access (SDMA). First, we consider a simple
scenario with two static ground nodes and analytically investigate the
effect of the UAV heading on the system sum rate.  We then study a
more general setting with multiple mobile ground-based terminals, and
develop an algorithm for dynamically adjusting the UAV heading in
order to maximize a lower bound on the ergodic sum rate of the uplink
channel, using a Kalman filter to track the positions of the mobile
ground nodes.  Fairness among the users can be guaranteed through
weighting the bound for each user's ergodic rate with a factor
inversely proportional to their average data rate.  For the common
scenario where a high $K$-factor channel exists between the ground
nodes and UAV, we use an asymptotic analysis to find simplified
versions of the algorithm for low and high SNR.  We present simulation
results that demonstrate the benefits of adapting the UAV heading in
order to optimize the uplink communications performance.  The
simulation results also show that the simplified algorithms perform
near-optimal performance.
\end{abstract}



%
\IEEEpeerreviewmaketitle

\newpage

\section{Introduction}
\subsection{Background}
There is increasing interest in the use of relatively small, flexible
unmanned aerial vehicles (UAVs) that fly at lower altitudes for
providing relay services for mobile {\em ad hoc} networks with
ground-based communication nodes
\cite{Cheng:2007, Han:2009, Freitas:2010, Palat:2005, Zhan:2011, Rubin:2007, Hillman:2002, Pinkney:1996}.  We
consider such an application in this paper, assuming a system with a
multi-antenna unmanned aerial vehicle (UAV) flying over a collection
of $N$ single-antenna mobile ground nodes. The UAV acts as a
decode-and-forward relay, sending the messages from the co-channel
users on the ground to some remote base station.  The goal is to
control the motion of the UAV so as to optimize the uplink
communications performance.

A number of different approaches have been proposed in the literature
to address the performance of UAV-assisted communication networks.
For example, in \cite{Cheng:2007}, a throughput maximization protocol
for non-real time applications was proposed for a network with UAV
relays in which the UAV first loads data from the source node and then
flies to the destination node to deliver it. The authors in
\cite{Han:2009} investigated different metrics for ad hoc network
connectivity and propose several approaches for improving the
connectivity through deployment of a UAV.  In \cite{Freitas:2010}, the
authors considered a scenario in which multiple UAVs are deployed to
relay data from isolated ground sensors to a base station, and an
algorithm was proposed to maintain the connectivity of the links
between the sensors and base station.

The work described above assumes that the ground nodes are static and
that the UAV is configured with only a single antenna.  Given the
well-known benefits of employing multiple antennas for communications,
it is natural to consider the advantages they offer for UAV-based
platforms\cite{Sharawi:2010}.  The measurement results of \cite{Kung:2010} showed that
using multiple receivers at the UAV can significantly increase the
packet delivery rate of the ground-to-air link.  A swarm of single
antenna UAVs were used as a virtual antenna array to relay data
from a fixed ad hoc network on the ground in \cite{Palat:2005}, and
the performance of distributed orthogonal space-time block codes
(OSTBC) and beamforming were evaluated.  A relay system with
multi-antenna UAVs and multi-antenna mobile ground terminals was
investigated in \cite{Zhan:2011}. The users employ OSTBC to transmit
data and the data transmissions are assumed to be interference
free. Based on estimates of the user terminals' future position, a
heading optimization approach was proposed that maximizes the uplink
sum rate of the network under the constraint that each user's rate is
above a given threshold.  The restriction of \cite{Zhan:2011} to the
interference-free case is a significant drawback, which we address in
this paper.  An earlier version of our work \cite{Jiang:2010}
discussed the use of an antenna array to improve the throughput of the
ground-to-air uplink when the users share the same channel and
interfere with one another.

\subsection{General Approach and Contribution}

In this paper, we consider a model similar to \cite{Zhan:2011}, with
several ground-based users communicating simultaneously with a
multi-antenna UAV.  The main difference with \cite{Zhan:2011} is that
we assume there exists co-channel interference between the different
users' data streams.  The users are assumed to transmit data with a
single antenna and the UAV uses beamforming to separate the co-channel
data streams.  We assume a correlated Rician fading channel model
between each ground node and the UAV, and we use a lower bound on the
ergodic achievable rate to quantify the uplink performance of the
relay network, assuming that the UAV uses a maximum
sig\-nal-to-int\-er\-fer\-ence-plus-noise ratio (SINR) beamformer for
interference mitigation. The strength of the mutual interference depends
on the correlation between the users' channel vectors, which in a high
$K$-factor channel is a function of the signals' angle of arrival
(AoA).  The AoAs depend in turn on the UAV's heading and the relative
positions of the UAV and the ground nodes.  Consequently, we propose
an adaptive algorithm for adjusting the heading of the UAV to minimize
the users' mutual interference and improve the uplink communications
performance.  In particular, the UAV is assumed to fly with a constant
velocity $v_u$, and it adjusts its heading in discrete time steps
(assuming a constraint on the maximum turn rate) in order to
optimize the bound on the achievable rate.  At time step $n$, the UAV
uses a Kalman filter driven by feedback from the ground terminals to
predict their positions at time $n+1$, and then the UAV computes its
heading in order to optimize the bound based on these future position
estimates.

The main results of this paper are summarized as follows:
\begin{enumerate}
\item We analyze the trajectory optimization problem for a special
case involving two static ground nodes. We use a rectangular-path
model to characterize the UAV's trajectory, which reduces the problem
to one of optimizing only the heading.  This problem can be solved
using a simple line search, and the results indicate how increasing
the size of the UAV array can reduce the system's sensitivity to the
heading direction.
\item For the case of a general network of mobile ground-based nodes,
we derive a lower bound on the average achievable sum rate to measure
the system performance. Based on this lower bound, we formulate a
heading optimization problem and propose a line-search algorithm to
adjust the UAV's heading direction at time step $n$ such that the
system performance at time step $n+1$ can be optimized.  We study the
performance of both time-division multiple access (TDMA) and
space-division multiple access (SDMA), and illustrate via simulation
the dramatic improvement offered by SDMA.
\item We derive asymptotic analytical results for the heading
optimization problem under the assumption of a high $K$-factor Rician
channel between the ground nodes and UAV.  The asymptotic results
provide simplified methods for solving the heading optimization
problem.  A separate approximation method is used for low and high SNR
cases, and we show that using the asymptotic expressions for 
heading optimization results in performance nearly identical to that of the
optimal algorithm.
\end{enumerate}

\subsection{Organization}
The organization of the paper is as follows. We present our assumed
signal and channel model in Section~\ref{sec:model}, and in
Section~\ref{sec:twouser} we focus on the UAV heading optimization
problem for the special case of two static ground users.  In
Section~\ref{sec:optnumerical}, we first describe the mobility model
for the UAV and ground nodes, as well as a standard Kalman filter for
predicting the future positions of the ground nodes. Then we formulate
the UAV heading optimization problem and propose an adaptive heading
adjustment algorithm.  We then derive asymptotic expressions for the
general heading optimization problem in Section~\ref{sec:asymptotic},
assuming a high-$K$ factor Rician channel between the UAV and ground
nodes.  Simulation results are provided in Section~\ref{sec:simu} to
illustrate the performance of the heading control algorithm, the
advantage of SDMA over TDMA, and the validity of the asymptotic
results.

\section{System Model}\label{sec:model}

\subsection{Signal Model}
We assume a UAV configured with an array of $M$ antennas, and a
collection of $N$ ground nodes each equipped with a single antenna.
We restrict attention to non-hovering UAVs that must maintain a
certain forward velocity to remain airborne.  We assume that, during
the period of time in which the UAV is receiving uplink data from the
ground nodes, the UAV maintains a constant altitude $h_u$ and a
constant velocity $v_u$.  For simplicity, we assume that each ground
node transmits with the same power $P_{t}$, but this assumption is
easily relaxed.  The signal received at the UAV array at time $n$ can
thus be written as
\begin{equation}
\mathbf{y}_n=\sum_{i=1}^{N}\sqrt{P_t}\mathbf{h}_{i,n}x_{i,n}+\mathbf{n}_n,
\end{equation}
where $\mathbf{h}_{i,n}\in\mathbb{C}^{M\times 1}$ is the channel
vector between node $i$ and the UAV, the data symbol $x_{i,n}$ is a
complex scalar with zero mean and unit magnitude,
$\mathbf{n}\in\mathbb{C}^{M\times 1}$ is zero-mean additive Gaussian
noise with covariance
$\mathbb{E}\{\mathbf{n}_n\mathbf{n}_n^{H}\}=\sigma^2\mathbf{I}_{M}$,
and $\mathbf{I}_{M}$ denotes an $M\times M$ identity matrix.  The UAV
isolates the data from the $i$th node by multiplying $\mathbf{y}_{n}$
with a beamformer $\mathbf{w}_{i,n}$.  As such, we assume that the
number of active uplink users is less than the number of antennas, or
$N \le M$.  Assuming the channels $\mathbf{h}_{i,n}, i=1,\dots, N$ are
known to the UAV ({\em e.g.,} via training data from the ground
nodes), the vector $\mathbf{w}_{i,n}$ that maximizes the
sig\-nal-to-int\-er\-fer\-ence-plus-noise ratio $SINR_{i,n}$ is given
by \cite{Winters:1984}
\begin{eqnarray}
\mathbf{w}_{i,n}=\mathbf{Q}_{i,n}^{-1}\mathbf{h}_{i,n}\; ,
\end{eqnarray}
where $\mathbf{Q}_{i,n}=\sum_{j=1,j\neq i}^{N}P_{t}\mathbf{h}_{j,n}\mathbf{h}_{j,n}^{H}+\sigma^2\mathbf{I}_{M}$. 
The corresponding $SINR_{i,n}$ can be calculated as
\begin{eqnarray}\label{eq:sinr}
SINR_{i,n}=P_t\mathbf{h}_{i,n}^{H}\mathbf{Q}_{i,n}^{-1}\mathbf{h}_{i,n}.
\end{eqnarray}

\subsection{Channel Model}
We assume a correlated Rician fading channel between each user node and the UAV
with consideration of large-scale path loss:
\begin{equation}\label{eq:dist}
\mathbf{h}_{i,n}=\frac{\mathbf{h}_{i,n}^{'}}{d_{i,n}^{\alpha}},
\end{equation}
where $\mathbf{h}_{i,n}^{'}$ is the normalized channel vector, $d_{i,n}$
is the distance between node $i$ and the UAV during the $n$th time step, and
$\alpha$ is the path loss exponent. Define the three dimensional
coordinates of the UAV and node $i$ as $(x_{u,n}, y_{u,n}, h_{u})$ and
$(x_{i,n}, y_{i,n}, 0)$, so that $d_{i,n}$ is given by
\begin{equation}
d_{i,n}=\sqrt{(x_{u,n}-x_{i,n})^2+(y_{u,n}-y_{i,n})^2+h_{u}^2}.
\end{equation}

For node $i$, we write the Rician fading channel vector
$\mathbf{h}_{i,n}^{'}$ with two components \cite{Bolcskei:2003}, a
line-of-sight (LOS) component $\bar{\mathbf{h}}_{i,n}$ and a Rayleigh
fading component $\tilde{\mathbf{h}}_{i,n}$:
\begin{equation}\label{eq:channel}
\mathbf{h}_{i,n}^{'}=\bar{\mathbf{h}}_{i,n}+\tilde{\mathbf{h}}_{i,n}.
\end{equation}
The LOS response will depend on the AoA of the signal, which in turn
depends on the heading of the UAV (determining the orientation of the
array) and the positions of the UAV and user nodes. For example,
assume a uniform linear array (ULA) with antennas separated by
one-half wavelength, and that at time step $n$ the phase delay
between adjacent antenna elements for the signal from the $i$th node
is $p_{i,n}$, then the LOS component could be modeled as
\begin{equation}\label{eq:ray}
\bar{\mathbf{h}}_{i,n}=\sqrt{\frac{K}{1+K}}\left[1, e^{jp_{i,n}},\cdots, e^{j(M-1)p_{i,n}}\right]^{T},
\end{equation}
where $K$ is the Rician $K$-factor. The phase delay $p_{i,n}$ is
calculated by\cite[chap. 4]{Varma:2002}
\begin{equation}
p_{i,n}=\pi\cos(\phi_{i,n})\sin(\theta_{i,n}),
\end{equation}
where $\phi_{i,n}$ and $\theta_{i,n}$ represent the elevation and
azimuth angles to the $i$th ground node.  In terms of the UAV and
user node positions, these quantities can be calculated as
\begin{eqnarray}\label{eq:aoa}
\cos(\phi_{i,n})&=&\sqrt{\frac{(x_{u,n}-x_{i,n})^2+(y_{u,n}-y_{i,n})^2}{(x_{u,n}-x_{i,n})^2+(y_{u,n}-y_{i,n})^2+h_u^2}},\nonumber\\
\sin(\theta_{i,n})&=&\cos(\delta_n-\epsilon_{i,n}),
\end{eqnarray}
where $\delta_n$ is the heading angle of the UAV, $\delta_n-\epsilon_{i,n}$ denotes
the angle between the UAV heading and the LOS to user $i$, and
\begin{eqnarray*}
\epsilon_{i,n} & = & \left\{\begin{array}{ll}
\zeta_{i,n},&y_{i,n}-y_{u,n}\ge 0\;\textrm{and}\;x_{i,n}-x_{u,n}\ge
0,\\ \zeta_{i,n}+\pi,&x_{i,n}-x_{u,n}\le 0,\\
\zeta_{i,n}+2\pi,&\textrm{otherwise}.
\end{array}\right. \\ 
\zeta_{i,n} & = & \arctan\left(\frac{y_{i,n}-y_{u,n}}{x_{i,n}-x_{u,n}}\right) \; .
\end{eqnarray*}
Since there is little multipath scattering near the UAV, any Rayleigh
fading components will experience high spatial correlation at the
receive end of the link.  Thus, we model the spatial correlated
Rayleigh component as
\begin{equation}
\tilde{\mathbf{h}}_{i,n}=\sqrt{\frac{1}{1+K}}(\mathbf{R}_{r})^{\frac{1}{2}}\mathbf{g}_{i,n},
\end{equation}
where $\mathbf{g}_{i,n}\in\mathbb{C}^{M\times 1}$ has
i.i.d. zero-mean, unit-variance complex Gaussian entries (which we
denote by $\mathcal{CN}(0,1)$), and $\mathbf{R}_{r}$ is the spatial
correlation matrix of the channel on the receiver side of the link. In
\cite{Tonu:1996}, a model for $\mathbf{R}_r$ is proposed under the
assumption that the multipah rays are distributed normally in two
dimensions around the angle from the source with standard deviation
$\sigma_{r}$, assuming a ULA receiver. We can easily extend this model
to take into account the third dimension corresponding to the
elevation angle, and the resulting $\mathbf{R}_{r}$ is given by
\begin{equation}\label{eq:Rr}
\mathbf{R}_r=\left(1+\frac{1}{K}\right)\bar{\mathbf{h}}_{i,n}
\bar{\mathbf{h}}_{i,n}^H\odot\mathbf{B}(\theta_{i,n},\sigma_{r}),
\end{equation}
where $\odot$ denotes the Hadamard (element-wise) product, and
\[
\mathbf{B}(\theta_{i,n},\sigma_{\phi})_{k,l}=e^{-\frac{1}{4}(\pi(k-l))^2\sigma_{r}^2\cos^2(\theta_{i,n})\left(1+\cos(2\phi_{i,n})-\frac{1}{2}\sigma_{r}^4\sin^2(2\phi_{i,n})(\pi(k-l))^2\cos^2(\theta_{i,n})\right)}\;. 
\]
The resulting distribution for $\mathbf{h}_{i,n}^{'}$ is thus
\begin{equation}\label{eq:distribution}
\mathbf{h}_{i,n}^{'}\thicksim\mathcal{CN}\bigg(\bar{\mathbf{h}}_{i,n}, \frac{1}{K+1}\mathbf{R}_{r}\bigg).
\end{equation} 

For the remainder of the paper, we will use the channel model defined
by the LOS component in~(\ref{eq:ray}) and the Rayleigh component
in~(\ref{eq:distribution}), which implies a UAV equipped with a ULA.
The ULA could be oriented along either the fuselage or the wings of
the UAV, the only difference being a $90^\circ$ change in how we
define the heading angle.  Extensions of the algorithm and analysis to
different array geometries would require one to use a different
expression for~(\ref{eq:ray}), which is straightforward, and to
derive a different spatial correlation matrix $\mathbf{R}_r$, which
is more complicated.

\section{Results for the Static Two-User Case}\label{sec:twouser}

To demonstrate the significant impact of the UAV trajectory on the
performance of the ground-to-air uplink, we first consider a simple
two user scenario.  The gross behavior of the UAV would be governed by
the distance $D$ between the two users, with three possibilities:
\begin{enumerate}
\item $D \gg h_u$ - This is not a particularly useful scenario for a
simultaneous uplink from both users since, if the UAV flies near their
midpoint, both users would experience low SINR at the UAV due to path
loss, and the sum data rate would be quite low.  In this case, a
better approach would likely involve the UAV serving each ground node
separately, circling directly above each user and alternately flying
between them.
\item $D \ll h_u$ - This case is also less interesting since the UAV
should obviously fly directly above the two users in as tight a
pattern as possible to minimize path loss.  The effect of the UAV
heading would be minimal, since the AoAs to the two users would be
nearly identical.  If the $K$-factor of the channel was high (as one
would expect when the UAV is essentially directly overhead), then the
channels would be highly correlated and a TDMA solution would likely
be preferred over SDMA.
\item $D = O(h_u)$ - Since the users transmit with the same power and
their channels have the same statistical properties, equalizing the
average uplink rates for the two users would require the UAV to fly a
symmetric trajectory centered around the midpoint of the two users.
If it was desired to minimize the variation in each user's average
uplink rate, the bounds of this trajectory would be small relative to
the distance to the users.  This is the case we consider in this
section.
\end{enumerate}

To make the analysis tractable, we focus on a rectangular trajectory
as depicted in Fig.~\ref{trajectory}, defined by the side lengths
$C_a$ and $C_b$ and the orientation $\delta$.  The angle $\delta$ is
defined to be with respect to the side of the rectangle with greater
length.  Given the assumptions for scenario~(3) above, the side
lengths are assumed to satisfy $\max\{C_a,C_b\} \le C_{\max} \ll
d_i$, so the figure is not to scale.  Under this assumption, the
performance of a rectangular trajectory is expected to be similar to
that for other trajectories with similar size and orientation ({\em
e.g.,} an ellipse or figure-8 pattern).  We also assume that
$\min\{C_a,C_b\} \ge C_{\min}$, which effectively accounts for the
turning radius of the UAV.  

The sum data rate at the UAV averaged along the trajectory is given by
\begin{eqnarray}
\bar{R}&=&\mathbb{E}\left\{\log_{2}(1+SINR_1)+\log_{2}(1+SINR_2)\right\}\nonumber\\
&=&\frac{1}{2(C_a+C_b)}\int_{\mathcal{C}}\left(\log_{2}(1+SINR_1(p))+\log_{2}(1+SINR_2(p))\right)dp,
\label{eq:Rbar}
\end{eqnarray}
where $\mathcal{C}$ denotes the rectangular path followed by the UAV,
variable $p$ denotes different positions along the trajectory and $dp$
represents the length of the elementary subintervals along the
trajectory.  The optimization problem we wish to solve is formulated as
\begin{eqnarray}\label{eq:trajecopt}
\max_{\delta,C_a,C_b}&&\bar{R}\\
\hspace{-1.5em}\mathrm{subject\quad to}&& 0 \le \delta \le \frac{\pi}{2}\nonumber\\
\hspace{-1.5em}&& C_{\min}\le C_b \le C_a \le C_{\max}\nonumber
\end{eqnarray}
where the symmetry of the problem allows us to restrict attention to
$0 \le \delta \le \pi/2$ and assume $C_b \le C_a$ without loss of
generality.  This non-linear optimization problem is difficult to
solve directly.  In the appendix, we show that for high SNR 
($\frac{P_t}{d_{i}^\alpha\sigma^2}\gg 1$) and assuming channels with
a large $K$-factor, the solution to~(\ref{eq:trajecopt}) is 
approximately given by $C_a=C_{\max}$, $C_b=C_{\min}$ and 
\begin{equation}\label{eq:optsimple}
\delta = \arg\min_{0 \le \delta \le \pi/2} \; \frac{R_c}{1+R_c}\frac{\sin^{2}(M\pi\cos(\phi^{'})
\cos(\delta))}{\sin^{2}(\pi\cos(\phi^{'})\cos(\delta))}+\frac{1}{1+R_c}
\frac{\sin^{2}(M\pi\cos(\phi^{'})\sin(\delta))}{\sin^{2}(\pi\cos(\phi^{'})\sin(\delta))},
\end{equation}
where $R_c=\frac{C_{\max}}{C_{\min}}$ and $\phi^{'}$ is the elevation angle to the two users at the center 
of the rectangle in Fig.~\ref{trajectory}, and satisfies
\[ \cos(\phi^{'}) =\frac{d_{i}}{\sqrt{d_{i}^2+h_u^2}}.\]
Minimizing~(\ref{eq:optsimple})
can be achieved by a simple line search over the interval $[0,\pi/2]$.

\begin{figure}
\centering
\includegraphics[height=3.5in, width=4.8in]{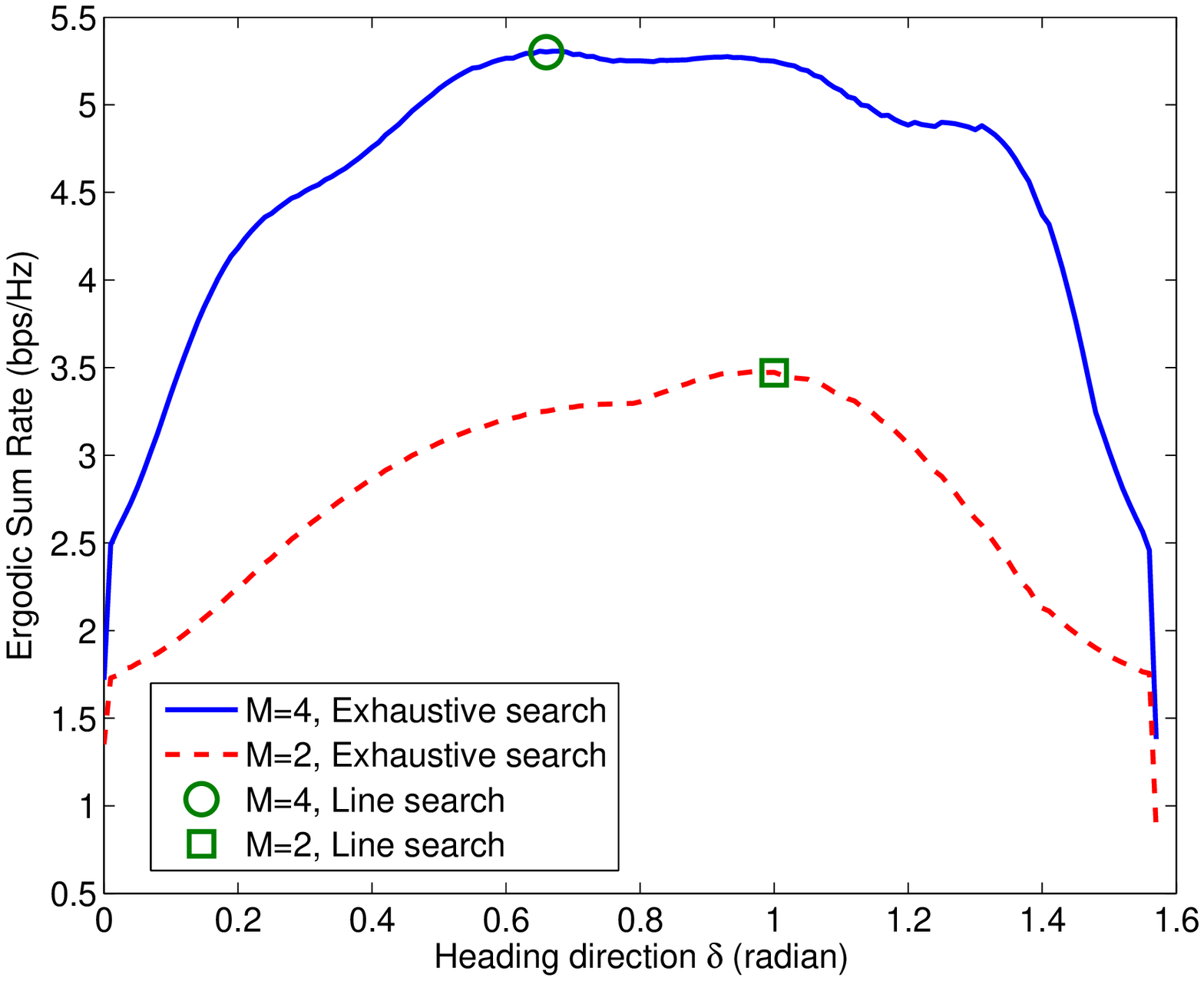}
\caption{Orientation of the rectangular trajectory provided by the
line search method in (\ref{eq:optsimple}). For the exhaustive search
method, the solid curve and the dashed curve denote the optimal sum
rate that can be achieved for different orientations $\delta$. When
$M=4$, the optimal $\delta$ are: 0.66\;(exhaustive search),
0.69\;(line search); when $M=2$, the optimal $\delta$ are:
0.98\;(exhaustive search), 1.00\;(line search).}\label{f1}
\end{figure}

To illustrate the validity of the approximate solution, we compare the
average system sum rate achieved by maximizing~(\ref{eq:trajecopt})
using an exhaustive search over $\{C_a,C_b\}$ for each value of
$\delta$ evaluated in the approximate line search
of~(\ref{eq:optsimple}).  The simulation parameters were
$d_1=d_2=1500\textrm{m}$, $h_u=350\textrm{m}$,
$C_{\min}=200\textrm{m}$, $C_{\max}=800\textrm{m}$, and
$\frac{P_t}{\sigma^2}=65\textrm{dB}$.  The results of the simulation
are plotted in Fig.~\ref{f1}, which shows the best rate obtained
by~(\ref{eq:trajecopt}) for each value of $\delta$, and the optimal
value obtained from minimizing~(\ref{eq:optsimple}) for $M=2$ and
$M=4$.  In both cases, the approximate approach
of~(\ref{eq:optsimple}) finds a trajectory orientation that yields a
near-optimal uplink rate.  Fig.~\ref{f1} also illustrates the benefit
of increasing the number of antennas at the UAV, and that proper
choice of the UAV heading can have a very large impact on
communications performance.
\section{Heading Optimization for a Mobile Ground Network}\label{sec:optnumerical}

In this section we consider a more general scenario in which several
mobile ground nodes are present and the UAV tracks their movement.  We
will consider both SDMA and TDMA approaches. In the SDMA scheme, all
of the ground nodes are transmitting simultaneously and the UAV uses
beamforming for source separation. For the TDMA method, each user is
allocated an equal time slot for data transmission.  It is assumed
that at time step $n-1$ all of the users feedback their current
position to the UAV, and these data are used to predict the positions
at time $n$.  An adaptive heading is proposed that calculates the UAV
heading at time step $n-1$ so that the network's performance at time
step $n$ will be optimized.

\subsection{Mobility Model and Position Prediction}\label{sec:kalman}

We adopt a first-order auto-regressive (AR) model for the dynamics of
the ground-based nodes \cite{Zaidi:2004}, and we assume the nodes
provide their location to the UAVs at each time step.  The UAV in
turn uses a Kalman filter to predict the positions of the nodes at the next
time step.  We define the dynamic state of user $i$ at time step $n-1$ as:
\begin{eqnarray}
\mathbf{s}_{i,n-1}=[x_{i,n-1}, y_{i,n-1}, v_{i,n-1}^{x}, v_{i,n-1}^{y}]^{T},
\end{eqnarray}
where $v_{i,n-1}^{x}$, $v_{i,n-1}^{y}$ denote the velocities in the
$x$ direction and $y$ direction respectively. According to the AR
model, the state of node $i$ at time step $n$ is given by
\begin{eqnarray}
\mathbf{s}_{i,n}&=&\mathbf{T}\mathbf{s}_{i,n-1}+\mathbf{w}_{i,n} \\
\mathbf{T}&=&\left[\begin{array}{cccc}
1&0&\Delta t&0\\
0&1&0&\Delta t\\
0&0&1&0\\
0&0&0&1\\
\end{array}\right] \; , 
\end{eqnarray}
where $\mathbf{w}_{i,n}\thicksim \mathcal{N}(0,
\sigma_{w}^2\mathbf{I}_{4})$ represents a process noise term.  Due to
the effects of delay, quantization and possible decoding errors, the 
UAV's knowledge of the ground nodes' position is imprecise.  This effect
is described by the measurement model for user $i$'s position:
\begin{eqnarray}
\mathbf{z}_{i,n}&=&\mathbf{F}\mathbf{s}_{i,n}+\mathbf{u}_{i,n} \\
\mathbf{F}&=&\left[\begin{array}{cccc}
1&0&0&0\\
0&1&0&0
\end{array}\right] \; ,
\end{eqnarray}
where
$\mathbf{u}_{i,n}\thicksim\mathcal{N}(0,\sigma_{u}^2\mathbf{I}_{2})$
represents the observation noise.  We assume a standard implementation
of the Kalman filter, as follows:

Initialization
\begin{equation}
\mathbf{x}_{i,0}=\mathbf{F}\mathbf{s}_{i,0},\;\mathbf{P}_{i,0}=\left[\begin{array}{cc} 0&0\\0&0\end{array}\right].
\end{equation}

Prediction
\begin{eqnarray}
\hat{\mathbf{s}}_{i,n|n-1}&=&\mathbf{T}\hat{\mathbf{s}}_{i,n-1|n-1},\\
\mathbf{P}_{i,n|n-1}&=&\mathbf{T}\mathbf{P}_{i,n-1|n-1}\mathbf{T}+\sigma_{w}^2\mathbf{I}_{4}.
\end{eqnarray}

Kalman gain
\begin{equation}
\mathbf{K}_{i,n}=\mathbf{P}_{i,n|n-1}\mathbf{F}^{T}(\mathbf{F}\mathbf{P}_{i,n|n-1}\mathbf{F}^{T}+\sigma_{u}^{2}\mathbf{I}_{2})^{-1}.
\end{equation}

Measurement update
\begin{eqnarray}
\hat{\mathbf{s}}_{i,n|n}&=&\hat{\mathbf{s}}_{i,n|n-1}+\mathbf{K}_{i,n}(\mathbf{z}_{i,n}-\mathbf{F}\hat{\mathbf{s}}_{i,n|n-1}),\\
\mathbf{P}_{i,n|n}&=&(\mathbf{I}_{4}-\mathbf{K}_{i,n}\mathbf{F})\mathbf{P}_{i,n|n-1}.
\end{eqnarray}

\subsection{SDMA Scenario}\label{sec:algorithm}

The average sum rate of the uplink network can be
approximated by a reasonably tight upper bound
\begin{eqnarray}\label{eq:sumrate}
C_{n}&=&\sum_{i=1}^{N}\mathbb{E}\left\{\log_{2}(1+SINR_{i,n})\right\}\nonumber\\
&\le&\sum_{i=1}^{N}\log_{2}\Big(1+\mathbb{E}\{SINR_{i,n}\}\Big).
\end{eqnarray}
The UAV heading $\delta_{n}$ will impact $C_{n}$ in two ways. First,
it will change the distance between the user nodes and the UAV during
time step $n$, which will impact the received power.  Second, and
often most importantly, changes in the heading will modify the AoA of
the LOS component, which impacts the ability of the beamformer to
spatially separate the users.  At time step $n-1$, based on the noisy
observation $\mathbf{z}_{i,n-1}$, the UAV uses the Kalman filter to
predict $(\hat{x}_{i,n}, \hat{y}_{i,n})$ and hence
$\mathbb{E}\{SINR_{i,n}\}$.  The heading optimization problem can thus
be formulated as
\begin{eqnarray}\label{eq:opt1}
\max_{\delta_{n}}&& \sum_{i=1}^{N} \log_2\big(1+\mathbb{E}\{SINR_{i,n}\}\big)\\
\mathrm{subject\quad to}&&  |\delta_{n}-\delta_{n-1}|\le\Delta\delta\nonumber \; ,
\end{eqnarray}
where $\Delta\delta$ represents that maximum change in UAV heading
possible for the given time step.  

The mean value of $SINR_{i,n}$ is calculated by
\begin{eqnarray}\label{eq:mean}
\mathbb{E}\{SINR_{i,n}\}&=&\mathbb{E}\big\{P_t\mathbf{h}_{i,n}^{H}\mathbb{E}\{\mathbf{Q}_{i,n}^{-1}\}\mathbf{h}_{i,n}\big\}\nonumber\\
&=&\frac{P_{t}}{d_{i,n}^{2\alpha}}\left(\frac{K}{K+1}\bar{\mathbf{h}}_{i,n}^{H}\mathbb{E}\{\mathbf{Q}_{i,n}^{-1}\}\bar{\mathbf{h}}_{i,n}+\frac{1}{K+1}\mathrm{tr}\Big(\mathbf{R}_{r}\mathbb{E}\{\mathbf{Q}_{i,n}^{-1}\}\Big)\right),
\end{eqnarray}
where $\mathrm{tr}(\cdot)$ denotes the trace operator.  Instead of 
working with the complicated term $\mathbb{E}\{\mathbf{Q}_{i,n}^{-1}\}$, 
we use instead the following lower bound based on Jensen's 
inequality\cite[Lemma 4]{zhang:2008}:
\begin{eqnarray}\label{eq:bound}
\mathbb{E}\{SINR_{i,n}\}&\ge&\frac{P_{t}}{d_{i,n}^{2\alpha}}\left(\frac{K}{K+1}\bar{\mathbf{h}}_{i,n}^{H}\mathbb{E}\{\mathbf{Q}_{i,n}\}^{-1}\bar{\mathbf{h}}_{i,n}+\frac{1}{K+1}\mathrm{tr}\Big(\mathbf{R}_{r}\mathbb{E}\{\mathbf{Q}_{i,n}\}^{-1}\Big)\right),
\end{eqnarray}
where
\[
\mathbb{E}\{\mathbf{Q}_{i,n}\}\!\!=\!\sum_{j=1,j\neq i}^{N}\!\frac{P_{t}}{d_{j,n}^{2\alpha}}
\big(\frac{K}{K+1}\bar{\mathbf{h}}_{j,n}\bar{\mathbf{h}}_{j,n}^{H}+
\frac{1}{K+1}\mathbf{R}_{r}\big)+\sigma^2\mathbf{I}_{M} \; .
\] 
We denote the lower bound on
the right side of equation (\ref{eq:bound}) as $\mathbb{E}_{l}\{SINR_{i,n}\}$
and substitute it into (\ref{eq:opt1}), leading to a related
optimization problem:
\begin{eqnarray}\label{eq:opt2}
\max_{\delta_{n}}&& \sum_{i=1}^{N} \log_2(1+\mathbb{E}_{l}\{SINR_{i,n}
\})\\
\mathrm{subject\quad to}&&  |\delta_{n}-\delta_{n-1}|\le\Delta\delta.\nonumber
\end{eqnarray}
Problem (\ref{eq:opt2}) requires finding the maximum value of a
single-variable function over a fixed interval $\delta_n \in
[\delta_{n-1}\!-\!\Delta\delta,\,\delta_{n-1}\!+\!\Delta\delta]$, and
thus can be efficiently solved using a one-dimensional line search.
Since problem (\ref{eq:opt2}) aims at maximizing the sum rate of the
system, the algorithm may lead to a large difference in achievable
rates between the users.  As an alternative, we may wish to guarantee
fairness among the users using, for example, the proportional fair
method~\cite{Holtzman:2001}:
\begin{eqnarray}\label{eq:opt4}
\max_{\delta_{n}}&& \sum_{i=1}^{N} w_{i,n}\log_2\big(1+\mathbb{E}_l\{SINR_{i,n}\}\big)\\
\mathrm{subject\quad to}&&  |\delta_{n}-\delta_{n-1}|\le\Delta\delta\nonumber,
\end{eqnarray} 
where $w_{i,n} \varpropto \bar{R}_{i,n}$ and $\bar{R}_{i,n}$ is 
user $i$'s average data rate:
\[
\bar{R}_{i,n}=\frac{1}{n-1}\sum_{k=1}^{n-1}\mathbb{E}\{\log_2(1+SINR_{i,k})\} \; .
\]

Based on our experience simulating the behavior of the algorithms   
described in~(\ref{eq:opt2}) and~(\ref{eq:opt4}), we propose two
simple refinements that eliminate undesirable UAV behavior.  First, to
avoid the UAV frequently flying back and forth between the user nodes
in an attempt to promote fairness, the weights $w_{i,n}$
in~(\ref{eq:opt4}) are only updated every $N_w$ time steps rather than
for every $n$.  Second, we expect that the optimal position of the UAV
should not stray too far from the center of gravity (CoG) of the
ground nodes.  This would not be the case if the users were clustered
into very widely separated groups, but such a scenario would likely
warrant the UAV serving the groups individually anyway.  To prevent
the UAV from straying too far from the CoG, at each time step the UAV
checks to see if the calculated heading would put it outside a certain
range $d_{\max}$ from the CoG.  If so, instead of using the calculated
value, it chooses a heading that points towards the CoG (or as close
to this heading as possible subject to the turning radius constraint).
Appropriate values for $N_w$ and $d_{\max}$ are found empirically.

The proposed adaptive heading algorithm is summarized in the 
following steps:
\begin{enumerate}
\item Use the Kalman filter to predict the user positions
$(\hat{x}_{i,n}, \hat{y}_{i,n})$ based on the noisy observations at
time step $n-1$, and construct the objective function
in~(\ref{eq:opt2}) or~(\ref{eq:opt4}) based on the predicted
positions.
\item Use a line search to find the solution of~(\ref{eq:opt2})
or~(\ref{eq:opt4}) for $\delta_{n}\in[0,2\pi]$, and denote the
solution as $\tilde{\delta_{n}}$. Calculate the heading interval
$\mathcal{O}_{n}=[\delta_{n-1}\!-\!\Delta\delta,
\delta_{n-1}\!+\!\Delta \delta]$. If
$\tilde{\delta}_{n}\in\mathcal{O}_{n}$, set
$\delta_{n}=\tilde{\delta_{n}}$, else set
$\delta_{n}=\arg\underset{\delta} \min |\delta-\tilde{\delta_{n}}|$,
where $\delta=\delta_{n-1}\!-\!\Delta\delta$ or
$\delta_{n-1}\!+\!\Delta \delta$.
\item Check to see if the calculated heading $\delta_n$ will place 
the UAV at a distance of $d_{\max}$ or greater from the predicted
CoG of the users.  If so, set $\delta_n=\delta_g$, where $\delta_g$
is the heading angle corresponding to the CoG, or set 
$\delta_{n}=\arg\underset{\delta} \min |\delta-\delta_g|$,
where $\delta=\delta_{n-1}\!-\!\Delta\delta$ or
$\delta_{n-1}\!+\!\Delta \delta$.
\item UAV flies with heading $\delta_n$ during time step $n$.
\end{enumerate}
Note that the line search in step 2 is over $[0,2\pi]$ rather than
just $[\delta_{n-1}\!-\!\Delta\delta,\,\delta_{n-1}\!+\!\Delta\delta]$, 
and the boundary point closest to the unconstrained maximum is chosen
rather than the boundary with the maximum predicted rate.  Thus, the
algorithm may temporarily choose a lower overall rate in pursuit of
the global optimum, although this scenario is uncommon.

\subsection{TDMA Scenario}
In the TDMA scenario, each node is assigned one time slot for
data transmission. After maximum ratio combining at the receiver,
the signal-to-noise ratio (SNR) of user $i$ is given by
\begin{eqnarray}
SNR_{i,n}=\frac{P_{t}}{\sigma^2}\|\mathbf{h}_{i,n}\|^2 , 
\end{eqnarray}
whose mean can be calculated as
\begin{eqnarray}
\mathbb{E}\{SNR_{i,n}\}=\frac{P_{t}M}{d_{i,n}^{2\alpha}\sigma^2}.
\end{eqnarray}
For the TDMA scenario, the optimization problem is formulated as
\begin{eqnarray}\label{eq:opt5}
\max_{\delta_{n}}&& \frac{1}{N}\sum_{i=1}^{N} w_{i,n}\log_2\bigg(1+\frac{P_{t}M}{d_{i,n}^{2\alpha}\sigma^2}\bigg)\\
\mathrm{subject\quad to}&&  |\delta_{n}-\delta_{n-1}|\le\Delta\delta\nonumber.
\end{eqnarray}
where 
\[ 
w_{i,n}=\left\{\rule{0mm}{8mm}\right.\begin{aligned}
&1 \quad\quad\textrm{max sum rate},\\
&\!\!\varpropto\bar{R}_{i} \;\,\textrm{proportional fair}.\\
\end{aligned}
\] 
The objective function in~(\ref{eq:opt5}) can be substituted in step 2
of the adaptive heading algorithm to implement the TDMA approach.

\section{Asymptotically Approximate Heading Algorithms}\label{sec:asymptotic}

Under certain conditions, we can eliminate the need for the bound
in~(\ref{eq:bound}) when defining our adaptive heading control
algorithm and simplify the algorithm implementation.  In this section,
we explore the asymptotic form of $SINR_{i,n}$ under both low and high
SNR conditions.  We show that in the low-SNR case, the optimal heading
can be found in closed-form, without the need for a line search.  In
the high-SNR case, we show that maximizing the sum rate is equivalent
to minimizing the sum of the users channel correlations, which can be
achieved by checking a finite set of candidate headings.  In
Section~\ref{sec:simu}, we show that the simpler asymptotic algorithms
derived here provide performance essentially identical to the
line-search algorithm of the previous section.  Our discussion here
will focus on the max-sum-rate case for SDMA; extensions to the
proportional fair and TDMA cases are straightforward.

\subsection{Asymptotic Analysis for Low SNR Case}

For low SNR $\frac{P_t}{d_{i,n}^{2\alpha}\sigma^2}\ll 1$, the average 
sum rate in~(\ref{eq:sumrate}) is approximated by
\begin{eqnarray}
C_{n}\approx\sum_{i}^{N}\mathbb{E}\{SINR_{i,n}\}
\end{eqnarray}
and problem~(\ref{eq:opt2}) can be rewritten as follows
\begin{eqnarray}\label{eq:optlow}
\max_{\delta_{n}}&&\sum_{i}^{N}\mathbb{E}\{SINR_{i,n}\}\\
\mathrm{subject\quad to}&&  |\delta_{n}-\delta_{n-1}|\le\Delta\delta.\nonumber
\end{eqnarray}
In this case we can approximate $\mathbf{Q}_{i,n}^{-1}$ with
the first order Neumann series \cite[Theorem 4.20]{Stewart:1998}:
\begin{equation}\label{eq:qapprox}
\mathbf{Q}_{i,n}^{-1}\approx\frac{1}{\sigma^2}\left(\mathbf{I}_{M}-\sum_{j=1,j\neq i}^{N}\frac{P_{t}}{\sigma^2}\mathbf{h}_{j,n}\mathbf{h}_{j,n}^{H}\right).
\end{equation}
Substituting (\ref{eq:qapprox}) into (\ref{eq:sinr}), the
$SINR_{i,n}$ for low SNR can be further expressed as
\begin{equation}
SINR_{i,n}=\frac{P_t}{\sigma^2}\mathbf{h}_{i,n}^{H}\left(\mathbf{I}_{M}-\sum_{j=1,j\neq i}^{N}\frac{P_{t}}{\sigma^2}\mathbf{h}_{j,n}\mathbf{h}_{j,n}^{H}\right)\mathbf{h}_{i,n},
\end{equation}
and we have
\begin{eqnarray}\label{eq:meansinr}
\mathbb{E}\left\{SINR_{i,n}\right\}&=&\mathbb{E}\left\{\frac{P_t}{\sigma^2}\mathbf{h}_{i,n}^{H}\left(\mathbf{I}_{M}-\sum_{j=1,j\neq i}^{N}\frac{P_{t}}{\sigma^2}\mathbf{h}_{j,n}\mathbf{h}_{j,n}^{H}\right)\mathbf{h}_{i,n}\right\}\nonumber\\
&=&\frac{P_{t}}{d_{i,n}^{2\alpha}\sigma^2}\left(\frac{K}{K+1}\bar{\mathbf{h}}_{i,n}^{H}\bigg(\mathbf{I}_{M}-\sum_{j=1,j\neq i}^{N}\frac{P_{t}}{d_{j,n}^{2\alpha}\sigma^2}\Big(\frac{K}{K+1}\bar{\mathbf{h}}_{j,n}\bar{\mathbf{h}}_{j,n}^{H}+\frac{1}{K+1}\mathbf{R}_r\Big)\bigg)\bar{\mathbf{h}}_{i,n}\right.\nonumber\\
&&\left.+\frac{1}{K+1}\mathrm{tr}\left(\mathbf{R}_{r}-\sum_{j=1,j\neq i}^{N}\frac{P_{t}}{d_{j,n}^{2\alpha}\sigma^2}\left(\frac{K}{K+1}\mathbf{R}_{r}\bar{\mathbf{h}}_{j,n}\bar{\mathbf{h}}_{j,n}^{H}+\frac{1}{K+1}\mathbf{R}_r^2\right)\right)\right)\nonumber\\
&\overset{(a)}{\approx}&\frac{P_{t}}{d_{i,n}^{2\alpha}\sigma^2}\left(M-\sum_{j=1,j\neq i}^{N}\frac{P_{t}}{d_{j,n}^{2\alpha}\sigma^2}|\bar{\mathbf{h}}_{i,n}^H\bar{\mathbf{h}}_{j,n}|^2\right),
\end{eqnarray}
where $(a)$ is based on the assumption of a large Rician factor $K$ for the
ground-to-air channel.  When scaled by
$\frac{P_{t}}{d_{i,n}^{2\alpha}\sigma^2}\ll 1$, the term involving
$|\bar{\mathbf{h}}_{i,n}^H\bar{\mathbf{h}}_{j,n}|^2$ in the above
equation plays a minor role in determining the value of
$\mathbb{E}\left\{SINR_{i,n}\right\}$. Assuming $\Delta\delta$ and the
ratio $\frac{v}{d_{i,n}}$ are small enough, we treat
$|\bar{\mathbf{h}}_{i,n}^H\bar{\mathbf{h}}_{j,n}|$ as a constant when
$\delta_n$ varies in
$[\delta_{n-1}-\Delta\delta,\delta_{n-1}+\Delta\delta]$. We
approximate $|\bar{\mathbf{h}}_{i,n}^H\bar{\mathbf{h}}_{j,n}|$ as
\begin{equation}\label{eq:correapprox}
|\bar{\mathbf{h}}_{i,n}^H\bar{\mathbf{h}}_{j,n}|\!\approx\!|\bar{\mathbf{h}}_{i,n}^{'H}\bar{\mathbf{h}}_{j,n}^{'}|\!=\!\left|\frac{\sin\Big(\frac{M\pi}{2}\big(\cos(\phi_{i,n}^{'})\cos(\delta_{n-1}\!-\!\epsilon_{i,n}^{'})\!-\!\cos(\phi_{j,n}^{'})\cos(\delta_{n-1}\!-\!\epsilon_{j,n}^{'})\big)\Big)}{\sin\Big(\frac{\pi}{2}\big(\cos(\phi_{i,n}^{'})\cos(\delta_{n-1}\!-\!\epsilon_{i,n}^{'})\!-\!\cos(\phi_{j,n}^{'})\cos(\delta_{n-1}\!-\!\epsilon_{j,n}^{'})\big)\Big)}\right|,
\end{equation}
where $\phi_{i,n}^{'}$ and $\epsilon_{i,n}^{'}$ are calculated assuming
the user nodes are located at $(\hat{x}_{i,n},\hat{y}_{i,n})$ and the
UAV is at $(x_{u,n-1},y_{u,n-1},h_u)$ with heading $\delta_{n-1}$. The
idea here is to use the UAV's position at
time step $n-1$ to calculate the users' AoA at time step
$n$. Moreover, $\frac{1}{d_{i,n}^{2\alpha}}$ can be approximated in
the following way
\begin{eqnarray}\label{eq:distance}
\frac{1}{d_{i,n}^{2\alpha}}&=&\Big((x_{u,n-1}+v\cos{\delta_n}-x_{i,n})^2+(y_{u,n-1}+v\sin{\delta_n}-y_{i,n})^2+h_{r}^2\Big)^{-\alpha}\nonumber\\
&=&\Big((x_{u,n-1}-x_{i,n})^2+(y_{u,n-1}-y_{i,n})^2+v^2+hr^2+2(x_{u,n-1}-x_{i,n})v\cos(\delta_n)\nonumber\\
&&+2(y_{u,n-1}-y_{i,n})v\sin(\delta_n)\Big)^{-\alpha}\nonumber\\
&\approx&a_{i,n}-b_{i,n}\cos(\delta_n)-c_{i,n}\sin(\delta_n),
\end{eqnarray}
where $a_{i,n}$, $b_{i,n}$ and $c_{i,n}$ are defined as follows
\begin{eqnarray}
a_{i,n}&=&\Big((x_{u,n-1}-x_{i,n})^2+(y_{u,n-1}-y_{i,n})^2+v^2+hr^2\Big)^{-\alpha} \nonumber\\
b_{i,n}&=&2\alpha v(x_{u,n-1}-x_{i,n})\Big((x_{u,n-1}-x_{i,n})^2+(y_{u,n-1}-y_{i,n})^2+v^2+hr^2\Big)^{-(\alpha+1)} \nonumber\\
c_{i,n}&=&2\alpha v(y_{u,n-1}-y_{i,n})\Big((x_{u,n-1}-x_{i,n})^2+(y_{u,n-1}-y_{i,n})^2+v^2+hr^2\Big)^{-(\alpha+1)} \; .\nonumber
\end{eqnarray}

Substituting~(\ref{eq:correapprox}) and~(\ref{eq:distance}) into (\ref{eq:meansinr}), 
$C_{n}$ can be approximated as
\begin{eqnarray}\label{eq:approx}
C_{n}&\approx&\frac{P_t}{\sigma^2}\!\sum_{i=1}^{N}\!M\!\left(a_{i,n}\!-\!b_{i,n}\cos(\delta_n)\!-\!c_{i,n}\sin(\delta_n)\right)\!-\!\!\left(\!\frac{P_t}{\sigma^2\!}\!\right)^2\!\sum_{i=1}^{N}\!\sum_{j=1,j\neq i}^{N}\!|\bar{\mathbf{h}}_{i,n}^{'H}\bar{\mathbf{h}}_{j,n}^{'}|^2\Big(\!a_{i,n}a_{j,n}\Big.\nonumber\\
&& \qquad \Big.\!-\!(a_{i,n}b_{j,n}\!+\!b_{i,n}a_{j,n})\cos(\delta_n)\!-\!(a_{i,n}c_{j,n}\!+\!c_{i,n}a_{j,n})\sin(\delta_n)\Big)\nonumber\\
&=&\frac{MP_t}{\sigma^2}\sum_{i=1}^{N}a_{i,n}\!-\!\left(\frac{P_t}{\sigma^2}\right)^2\sum_{i=1}^{N}\sum_{j=1,j\neq i}^{N}\!\!|\bar{\mathbf{h}}_{i,n}^{'H}\bar{\mathbf{h}}_{j,n}^{'}|^2a_{i,n}a_{j,n}\!-\!\left(\frac{MP_t}{\sigma^2}\sum_{i=1}^{N}b_{i,n}\right.\nonumber\\
&&\left.\!-\!\left(\frac{P_t}{\sigma^2}\right)^2\sum_{i=1}^{N}\sum_{j=1,j\neq i}^{N}\!\!|\bar{\mathbf{h}}_{i,n}^{'H}\bar{\mathbf{h}}_{j,n}^{'}|^2(a_{i,n}b_{j,n}+b_{i,n}a_{j,n})\right)\!\cos(\delta_n)\!-\!\left(\frac{MP_t}{\sigma^2}\sum_{i=1}^{N}c_{i,n}\right.\nonumber\\
&&\left.-\left(\frac{P_t}{\sigma^2}\right)^2\sum_{i=1}^{N}\sum_{j=1,j\neq i}^{N}\!\!|\bar{\mathbf{h}}_{i,n}^{'H}\bar{\mathbf{h}}_{j,n}^{'}|^2(a_{i,n}c_{j,n}+c_{i,n}a_{j,n})\right)\!\sin(\delta_n).
\end{eqnarray}
Define the first two terms in (\ref{eq:approx}) as $A_{n}$, and the
term multiplying $\cos(\delta_n)$ and $\sin(\delta_n)$ as $B_{n}$ and
$D_{n}$, respectively. Then (\ref{eq:approx}) can be further expressed as
\begin{equation}
C_{n}=A_{n}-\sqrt{B_{n}^2+D_{n}^2}\cos(\delta_n-\psi_n),
\end{equation}
where 
\[
\psi_n=\left\{\rule{0mm}{8mm}\right.\begin{aligned}
&\arctan\left(\frac{D_{n}}{B_{n}}\right) \qquad \mathrm{if}\;B_{n}\ge 0,\\
&\arctan\left(\frac{D_{n}}{B_{n}}\right)+\pi \qquad \mathrm{otherwise}.\\
\end{aligned} 
\] 
From this expression, we see that the average sum rate $C_{n}$ can be
written as a sinusoidal function of $\delta_n$, and the maximizing
heading $\delta_n$ is given by
\begin{equation}
\delta_{n}^{*}=\bmod_{2\pi}(\psi_{n}+\pi).
\end{equation}
As a result, for low-SNR, the following closed-form approximation
to problem~(\ref{eq:optlow}) can be used:
\begin{eqnarray}
\delta_n=\left\{\begin{array}{ll}
\delta_n^{*} & \delta_{n-1}-\Delta\delta<\delta_n^{*}<\delta_{n-1}+\Delta\delta\\
\delta_{n-1}-\Delta\delta & \bmod_{\pi}(|\delta_{n-1}-\Delta\delta-\delta_n^{*}|)
<\bmod_{\pi}(|\delta_{n-1}+\Delta\delta-\delta_n^{*}|)\\
\delta_{n-1}+\Delta\delta & \bmod_{\pi}(|\delta_{n-1}-\Delta\delta-\delta_n^{*}|)\ge
\bmod_{\pi}(|\delta_{n-1}+\Delta\delta-\delta_n^{*}|) \; .\\
\end{array}\right.
\end{eqnarray}

\subsection{Asymptotic Analysis for High SNR Case}

In the high SNR case where $\frac{P_t}{d_{i,n}^{2\alpha}\sigma^2}\gg
1$, the average sum rate maximization problem can be approximated as
\begin{eqnarray}\label{eq:opthigh}
\max_{\delta_{n}}&&\prod_{i=1}^{N}\mathbb{E}\{SINR_{i,n}\}\\
\mathrm{subject\quad to}&&  |\delta_{n}-\delta_{n-1}|\le\Delta\delta.\nonumber
\end{eqnarray}
Here, when $\frac{P_t}{d_{i,n}^{2\alpha}\sigma^2}\gg 1$, we
approximate $\mathbf{Q}_{i,n}^{-1}$ as follows:
\begin{eqnarray}\label{eq:qapprox2}
\mathbf{Q}_{i,n}^{-1}&=&\frac{1}{\sigma^2}\left(\mathbf{I}_{M}+\frac{P_t}{\sigma^2}\mathbf{H}_{i,n}\mathbf{D}_{i,n}\mathbf{H}_{i,n}^H\right)^{-1}\nonumber\\
&\overset{(b)}{=}&\frac{1}{\sigma^2}\left(\mathbf{I}_{M}-\frac{P_t}{\sigma^2}\mathbf{H}_{i,n}\mathbf{D}_{i,n}\left(\mathbf{I}_{M}+\frac{P_t}{\sigma^2}\mathbf{H}_{i,n}^H\mathbf{H}_{i,n}\mathbf{D}_{i,n}\right)^{-1}\mathbf{H}_{i,n}^H\right)\nonumber\\
&\overset{(c)}{\approx}&\frac{1}{\sigma^2}\left(\mathbf{I}_{M}-{\mathbf{H}}_{i,n}\big(\mathbf{H}_{i,n}^H\mathbf{H}_{i,n}\big)^{-1}\mathbf{H}_{i,n}^H\right),
\end{eqnarray}
where (b) is due to the matrix inversion lemma, (c) is due to the
approximation $\left(\mathbf{I}_{M}+\frac{P_t}{\sigma^2}\mathbf{H}_{i,n}^H\mathbf{H}_{i,n}
\mathbf{D}_{i,n}\right)^{-1}\approx\left(\frac{P_t}{\sigma^2}\mathbf{H}_{i,n}^H
\mathbf{H}_{i,n}\mathbf{D}_{i,n}\right)^{-1}$, and
\begin{eqnarray*}
\mathbf{D}_{i,n}& = & \mathrm{diag}\left\{\frac{1}{d_{1,n}^{2\alpha}}, \cdots, 
\frac{1}{d_{i-1,n}^{2\alpha}},\frac{1}{d_{i+1,n}^{2\alpha}},\cdots,\frac{1}{d_{N,n}^{2\alpha}}\right\} \\
\mathbf{H}_{i,n} & = & [\mathbf{h}_{1,n} \; \cdots \; \mathbf{h}_{i-1,n} \; \mathbf{h}_{i+1,n} \; \cdots \; \mathbf{h}_{N,n}]
\end{eqnarray*}
are formed by eliminating the terms for user $i$.
Plugging~(\ref{eq:qapprox2}) into~(\ref{eq:sinr}), we obtain
\begin{equation}
SINR_{i,n}\approx\frac{P_t}{\sigma^2d_{i,n}^{2\alpha}}\left(\mathbf{h}_{i,n}^{H}\mathbf{h}_{i,n}-\big\|\mathbf{h}_{i,n}^H\mathbf{H}_{i,n}\big(\mathbf{H}_{i,n}^H\mathbf{H}_{i,n}\big)^{-1}\mathbf{H}_{i,n}^H\big\|^2\right),
\end{equation}
For large $K$-factor channels we ignore the contribution of the
Rayleigh term, so that 
\begin{equation}
\mathbb{E}\{SINR_{i,n}\}\approx\frac{P_t}{\sigma^2d_{i,n}^{2\alpha}}\left(M-\big\|\bar{\mathbf{h}}_{i,n}^H\bar{\mathbf{H}}_{i,n}\big(\bar{\mathbf{H}}_{i,n}^H\bar{\mathbf{H}}_{i,n}\big)^{-1}\bar{\mathbf{H}}_{i,n}^H\big\|^2\right),
\end{equation}
where $\bar{\mathbf{H}}_{i,n}$ is defined similarly to $\mathbf{H}_{i,n}$.
Thus, the heading optimization problem can be written as
\begin{eqnarray}\label{eq:opthigh2}
\max_{\delta_{n}}&&\prod_{i=1}^{N}\frac{P_t}{d_{i,n}^{\alpha}\sigma^2}\prod_{i=1}^N \left(M-\big\|\bar{\mathbf{h}}_{i,n}^H\bar{\mathbf{H}}_{i,n}\big(\bar{\mathbf{H}}_{i,n}^H\bar{\mathbf{H}}_{i,n}\big)^{-1}\bar{\mathbf{H}}_{i,n}^H\big\|^2\right)\\
\mathrm{subject\quad to}&&  |\delta_{n}-\delta_{n-1}|\le\Delta\delta.\nonumber
\end{eqnarray}

At this point we make two further approximations.  First, we will
ignore the terms in the product involving $1/d_{i,n}$, since $d_{i,n}$
will not change appreciably over one time step compared with the terms
involving products of $\bar{\mathbf{h}}_{i,n}$, which are
angle-dependent.  Second, we will make the assumption that the matrix
$\bar{\mathbf{H}}_{i,n}^H \bar{\mathbf{H}}_{i,n}$ is approximately
diagonal, which implies that the UAV attempts to orient itself so that
the correlation between the mean channel vectors for different users
is minimized.  If we then apply these two assumptions
to~(\ref{eq:opthigh2}), we find that the heading problem reduces to
\begin{eqnarray}{\label{eq:approxopt3}}
\min_{\delta_{n}}&&\sum_{i=1}^{N}\sum_{j=i+1}^{N}|\bar{\mathbf{h}}_{i,n}^H\bar{\mathbf{h}}_{j,n}|\\
\hspace{-1.5em}\mathrm{subject\quad to}&&  |\delta_{n}-\delta_{n-1}|\le\Delta\delta \; ,\nonumber
\end{eqnarray}
which is consistent with the assumption of minimizing inter-user
channel correlation.

\begin{figure}
\centering
\includegraphics[height=3.5in, width=4.8in]{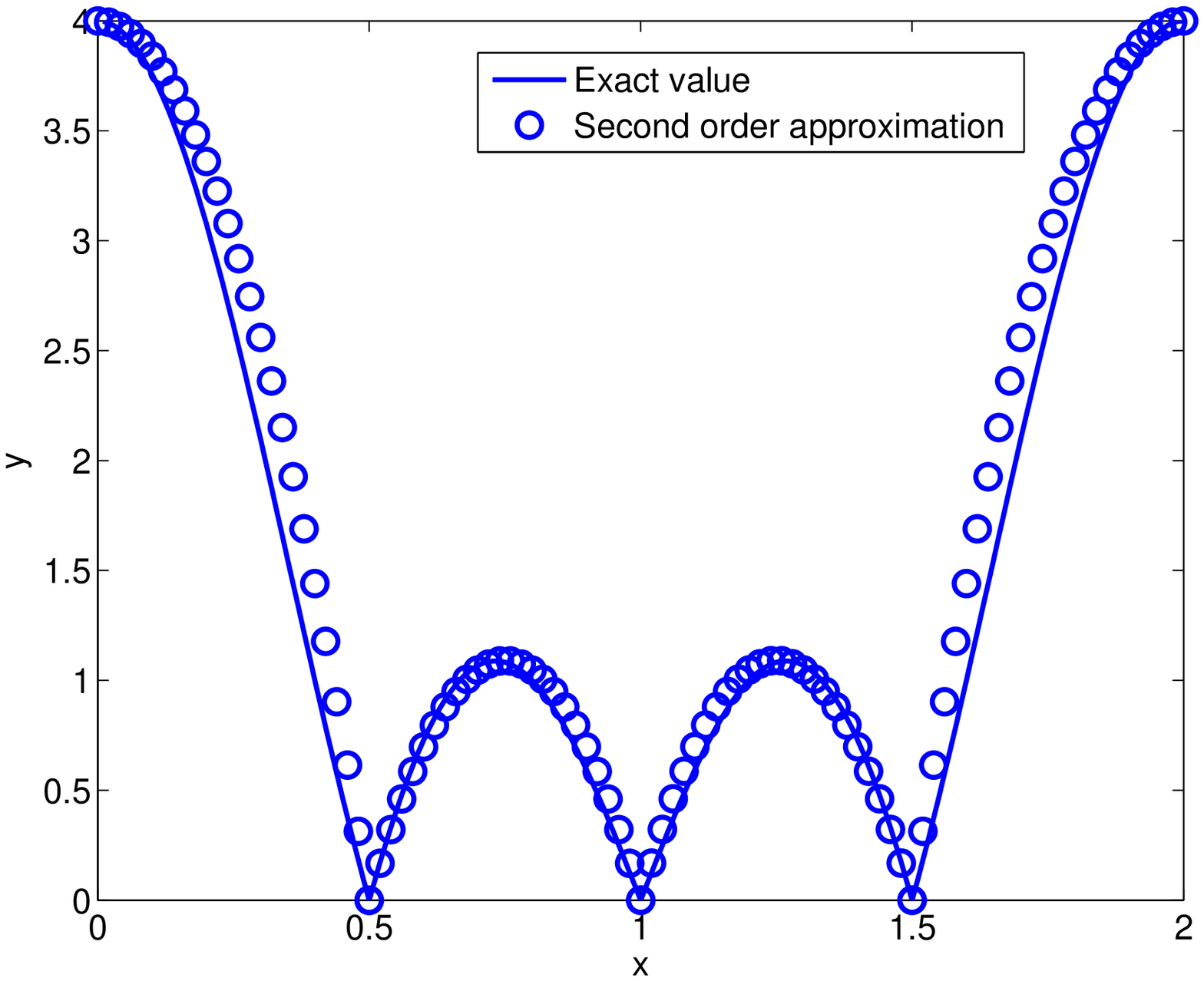}
\caption{Plot of $|\bar{\mathbf{h}}_{i}^H\bar{\mathbf{h}}_j|$ as a function of the AoA between the two users, along with a set of piecewise quadratic approximations.}\label{f2}
\end{figure}

In Fig.~\ref{f2}, we show a plot of
$|\bar{\mathbf{h}}_{i,n}^{H}\bar{\mathbf{h}}_{j,n}|$ for $M=4$ as a
function of the difference in AoA between the two users (variable $x$
in the plot).  It is clear that
$|\bar{\mathbf{h}}_{i,n}^{H}\bar{\mathbf{h}}_{j,n}|$ is a piecewise
concave function.  Since a sum of concave functions is also concave,
the criterion in~(\ref{eq:approxopt3}) is piecewise concave as well.
Since the minimum of a concave function must be located at the
boundary of its domain, to find the solution to~(\ref{eq:approxopt3})
it is enough to evaluate the criterion at the boundary points
$\{\delta_{n-1}-\Delta\delta,\delta_{n-1}+\Delta\delta\}$ and the zero
points of $|\bar{\mathbf{h}}_{i,n}^H\bar{\mathbf{h}}_{j,n}|$ located
within $[\delta_{n-1}-\Delta\delta,\delta_{n-1}+\Delta\delta]$.
To find the zero locations, we use the fact that a piecewise quadratic
approximation to $|\bar{\mathbf{h}}_{i,n}^{H}\bar{\mathbf{h}}_{j,n}|$ 
is very accurate (as depicted in Fig.~\ref{f1}).  When $\Delta\delta$
is not too large, the phase term $p_{i,n}$ in~(\ref{eq:ray}) satisfies
\begin{eqnarray}\label{eq:angelapp}
p_{i,n}&\approx&\pi\cos(\phi_{i,n}^{'})\left(\cos(\epsilon_{i,n}^{'}-\delta_{n-1})+\sin(\epsilon_{i,n}^{'}-\delta_{n-1})(\delta_n-\delta_{n-1})\right)=e_{i,n}+f_{i,n}x,
\end{eqnarray}
where $x=\delta_{n}-\delta_{n-1}$, $e_{i,n}=\pi\cos(\phi_{i,n}^{'})\cos(\epsilon_{i,n}^{'}-\delta_{n-1})$, $f_{i,n}=\pi\cos(\phi_{i,n}^{'})\sin(\epsilon_{i,n}^{'}-\delta_{n-1})$, $x\in [-\Delta\delta,\Delta\delta]$
and the calculation of $\phi_{i,n}^{'}$ and $\epsilon_{i,n}^{'}$ 
follows~(\ref{eq:correapprox}).
Based on (\ref{eq:angelapp}), we obtain
\begin{equation}
|\bar{\mathbf{h}}_{i,n}^{H}\bar{\mathbf{h}}_{j,n}|\approx\left|\frac{\sin\Big(\frac{M}{2}\big((f_{i,n}-f_{j,n})x+e_{i,n}-e_{j,n}\big)\Big)}{\sin\Big(\frac{1}{2}\big((f_{i,n}-f_{j,n})x+e_{i,n}-e_{j,n}\big)\Big)}\right|.
\end{equation}
Then the zero points of
$|\bar{\mathbf{h}}_{i,n}^{H}\bar{\mathbf{h}}_{j,n}|$ in terms of $x$
are approximately given by\footnote{Where we assume $\Delta\delta<1$,
$\left|\left(f_{i,n}-f_{j,n}\right)x+e_{i,n}-e_{j,n}\right|<4\pi$ and
we only consider the zero points in $[-4\pi,4\pi]$.}
\begin{equation}
z_{k}^{i,j}=\frac{2k\pi/M-e_{i,n}+e_{j,n}}{f_{i,n}-f_{j,n}}, \qquad k=\pm 1,\dots,\pm 2M-1.
\end{equation}
Finally, the asymptotic solution to problem (\ref{eq:approxopt3}) can
be written as
\begin{equation}
\delta_n=\arg\min_{\delta_n} \sum_{i=1}^{N}\sum_{j=i+1}^{N}|\bar{\mathbf{h}}_{i,n}^{H}\bar{\mathbf{h}}_{j,n}|,\; \delta_n\in\{z_{k}^{i,j}\in[-\Delta\delta,\Delta\delta]\}\cup\{\delta_{n-1}-\Delta\delta,\delta_{n-1}+\Delta\delta\}.
\end{equation}

\section{Simulation Results}\label{sec:simu}

A simulation example involving a UAV with a 4-element ULA and four
user nodes was carried out to test the performance of the proposed
algorithm. The time between UAV heading updates was set to $\Delta
t=1\textrm{s}$, and the simulation was conducted over $L=300$
steps. In the simulation, we assume the users have the same initial
velocity, and then move independently according to the model described
earlier. The initial velocity of the nodes is $10\textrm{m/s}$, and
their initial positions in meters are $(0, 25)$, $(240, 20)$, $(610,
30)$, $(1240, 20)$. The elements of the process and measurement noise vectors are
assumed to be independent with variances given by $\sigma_{w}^{2}=0.5$
and $\sigma_{u}^{2}=0.1$, respectively. The user's transmit power is
set to $\frac{P_{t}}{\sigma^2}=45\textrm{dB}$ and the path loss
exponent is $\alpha=1$. When $L=150$, all the nodes make a sharp turn
and change their velocity according to
${v^{y}_{i,150}}/{v^{x}_{i,150}}=-1.8856$. The initial position of the
UAV is $(x_{u,0}, y_{u,0})=(50, 100)\textrm{m}$ and its altitude is
assumed to be $h_{u}=350\textrm{m}$. The speed of the UAV is
$v_{u}=50\textrm{m/s}$, and the maximum heading angle change is set to
be either $\Delta\delta=\frac{\pi}{6}$ or $\frac{\pi}{9}$ depending on
the case considered.  The angle spread factor in~(\ref{eq:Rr}) is set
to $\sigma_{r}^2=0.05$. For the proportional fair case, $N_w$ is set
to $4$ and for the high SNR case, $d_{\max}$ is set to
$300\textrm{m}$.

\begin{figure}
\centering
\includegraphics[height=3.5in, width=4.8in]{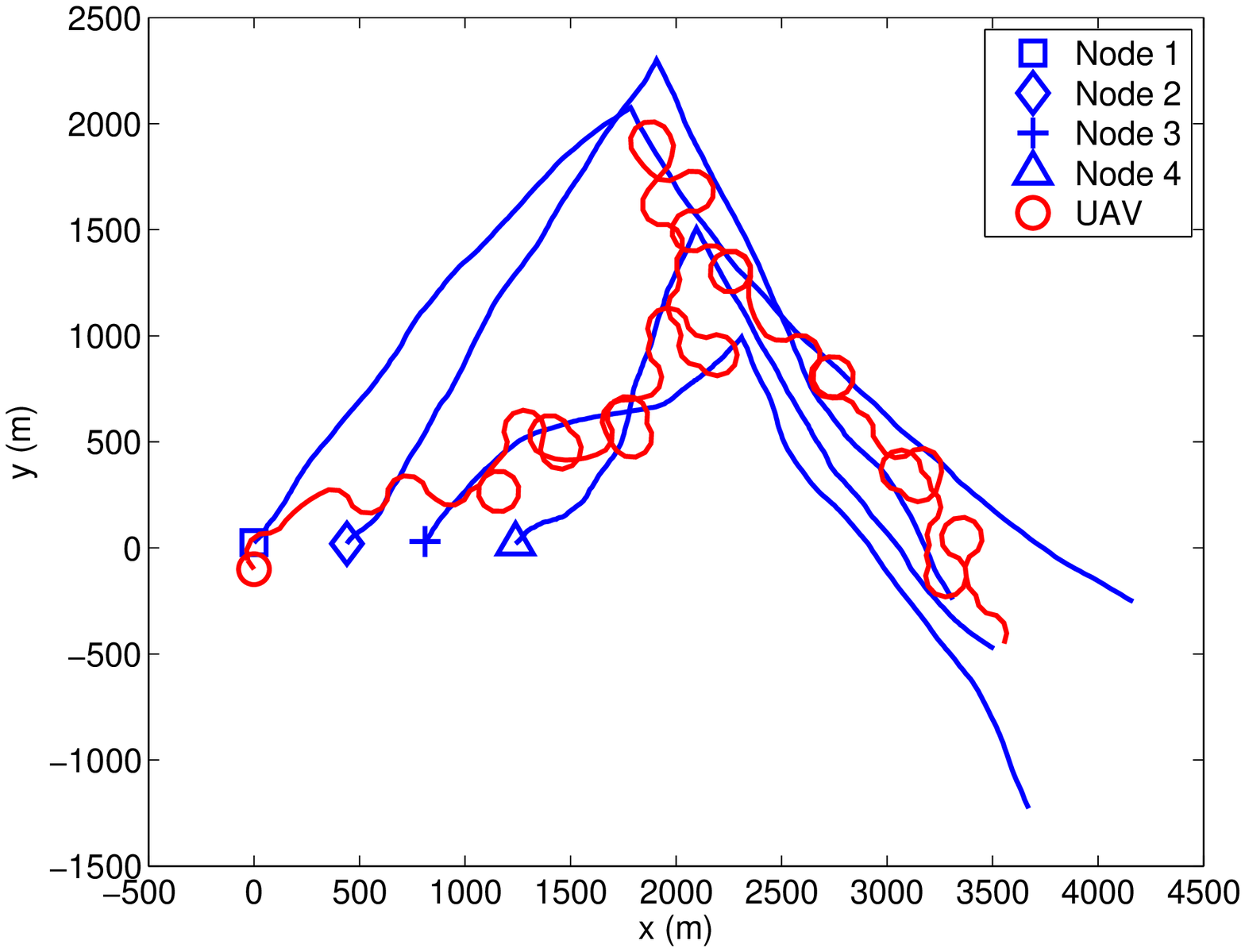}
\caption{Trajectories of the UAV and user nodes for SDMA with $\Delta\delta=\frac{\pi}{6}$, $K=10$ and $\frac{P_t}{\sigma^2}=45\textrm{dB}$, maximizing sum rate. The average sum rate is: $1.8185\textrm{bps/Hz}$. The single user data rates are $\textrm{u}_1=0.5607, \textrm{u}_2=0.6138, \textrm{u}_3=0.2406, \textrm{u}_4=0.4034$.}\label{f3}
\end{figure}

\begin{figure}
\centering
\includegraphics[height=3.5in, width=4.8in]{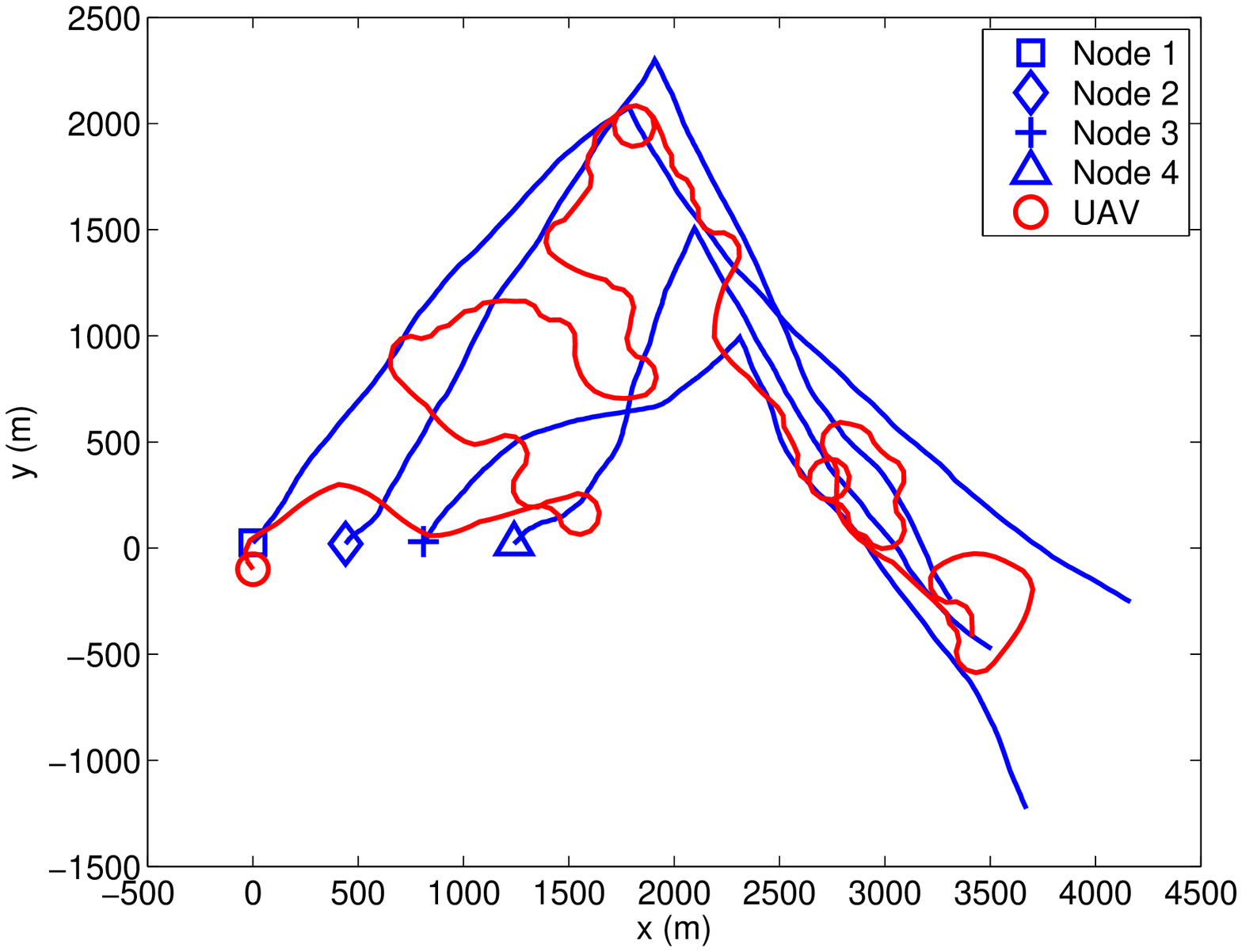}
\caption{Trajectories of the UAV and user nodes for SDMA with $\Delta\delta=\frac{\pi}{6}$, $K=10$ and $\frac{P_t}{\sigma^2}=45\textrm{dB}$, proportional fair. The average sum rate is $1.6968\textrm{bps/Hz}\;(\textrm{u}_1=0.4169, \textrm{u}_2=0.4084, \textrm{u}_3= 0.4088, \textrm{u}_4=0.4627$).}\label{f4}
\end{figure}

\begin{figure}
\centering
\includegraphics[height=3.5in, width=4.8in]{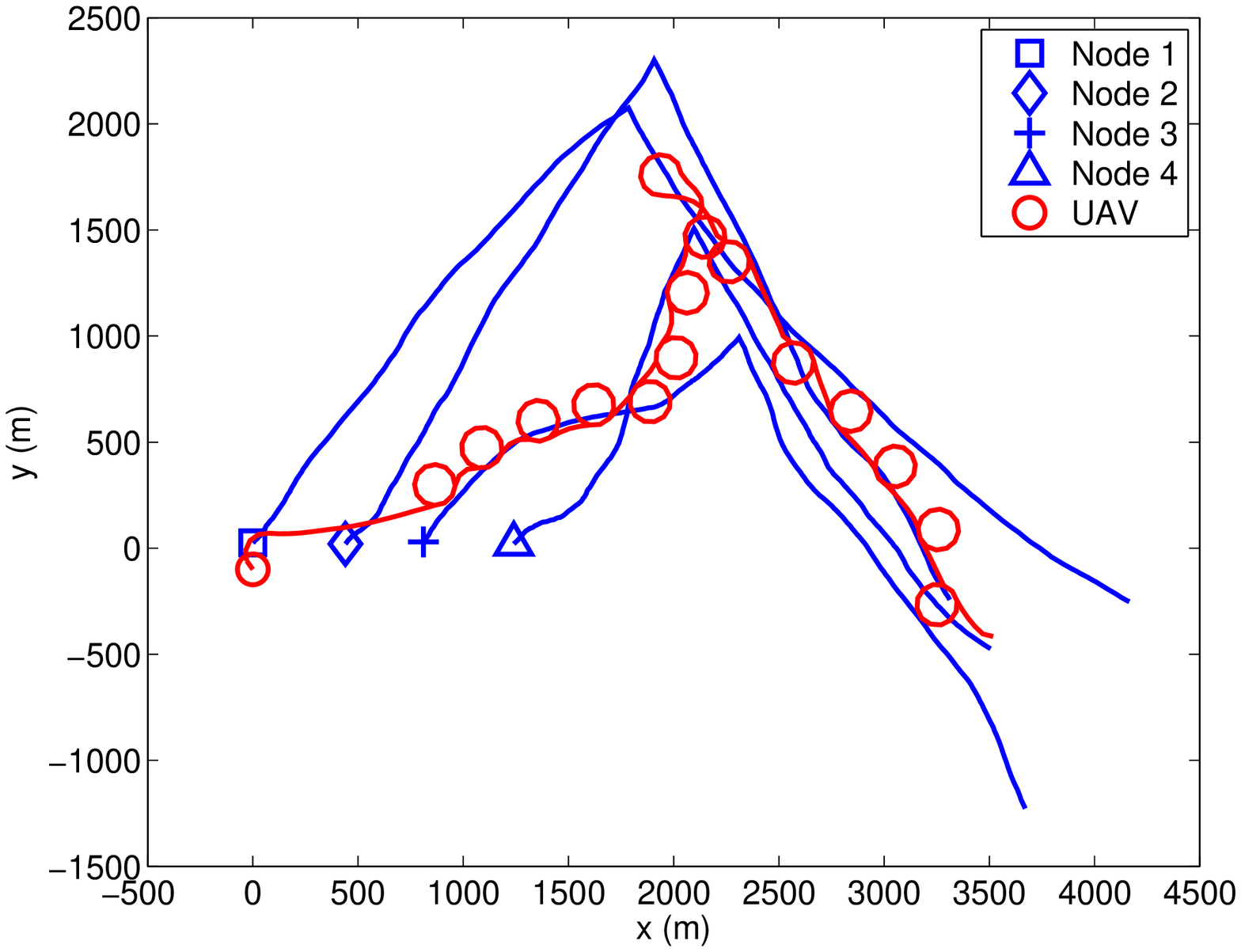}
\caption{Trajectories of the UAV and user nodes for TDMA with $\Delta\delta=\frac{\pi}{6}$, $K=10$ and $\frac{P_t}{\sigma^2}=45\textrm{dB}$, maximizing sum rate. The average sum rate is: $0.5294\textrm{bps/Hz}\;(\textrm{u}_1=0.1418, \textrm{u}_2=0.1674, \textrm{u}_3=0.0895, \textrm{u}_4=0.1307)$.}\label{f5}
\end{figure}

\begin{figure}
\centering
\includegraphics[height=3.5in, width=4.8in]{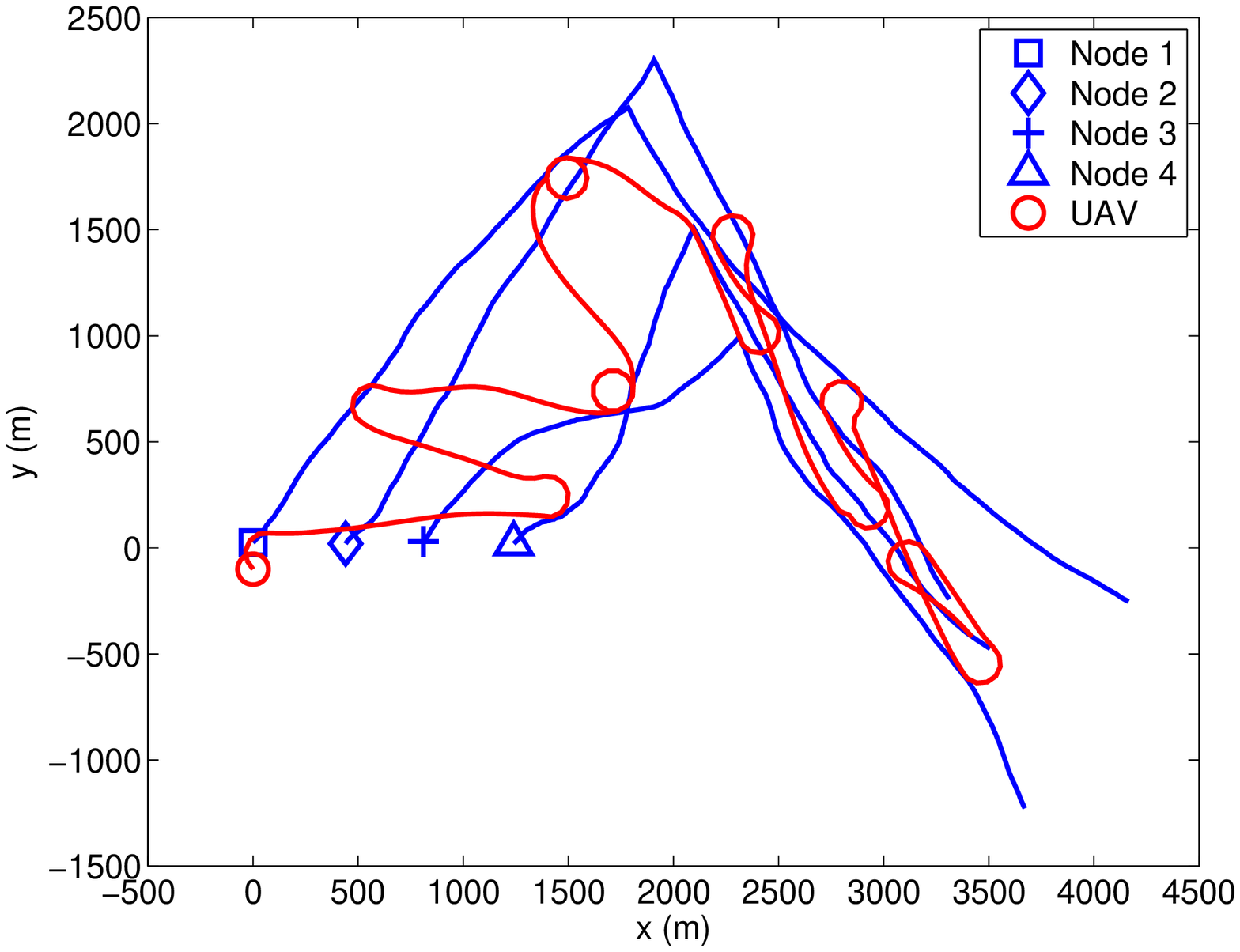}
\caption{Trajectories of the UAV and user nodes for TDMA with $\Delta\delta=\frac{\pi}{6}$, $K=10$ and $\frac{P_t}{\sigma^2}=45\textrm{dB}$, proportional fair. The average sum rate is: $0.5139\textrm{bps/Hz}\;(\textrm{u}_1=0.1222, \textrm{u}_2=0.1274, \textrm{u}_3=0.1193, \textrm{u}_4=0.1450$).}\label{f6}
\end{figure}

Figs.~\ref{f3}-\ref{f6} show the trajectories of the UAV and mobile
nodes for the SDMA and the TDMA scenarios assuming either max-sum or
proportional fair objective functions and
$\Delta\delta=\frac{\pi}{6}$.  The decision-making behavior of the UAV
is evident from its ability to appropriately track the nodes as they
dynamically change position.  Due to the relatively high speed of the
UAV, loop maneuvers are necessary to maintain an optimal position for
the uplink communications signals.  In the proportional-fair approach,
the UAV tends to visit the nodes in turn, while the max-sum rate
algorithm leads to the UAV approximately tracking the area where the
user node density is highest.  Note that in this example the
proportional-fair algorithm only suffers a slight degradation in
overall sum rate compared with the max-sum rate approach.

\begin{figure}
\centering
\includegraphics[height=3.4in, width=4.6in]{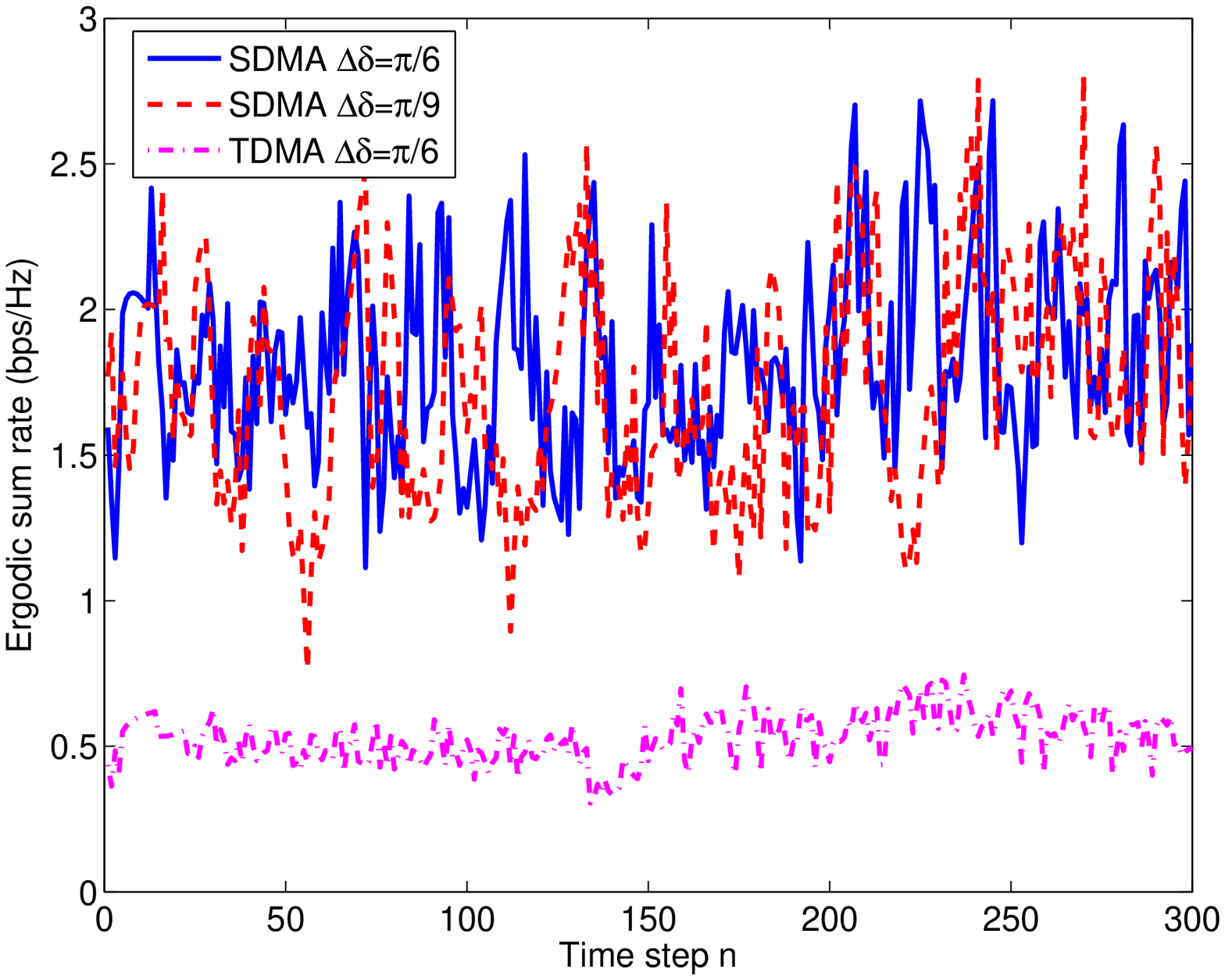}
\caption{Comparison of sum rate performance (bps/Hz) with $K=10$ and $\frac{P_t}{\sigma^2}=45\textrm{dB}$, maximizing sum rate. The average sum rates are: $1.8185\;(\textrm{SDMA}, \Delta\delta=\frac{\pi}{6})$, $1.7247\;(\textrm{SDMA}, \Delta\delta=\frac{\pi}{9})$, $0.5294\;(\textrm{TDMA}, \Delta\delta=\frac{\pi}{6})$.}\label{f7}
\end{figure}

\begin{figure}
\centering
\includegraphics[height=3.4in, width=4.6in]{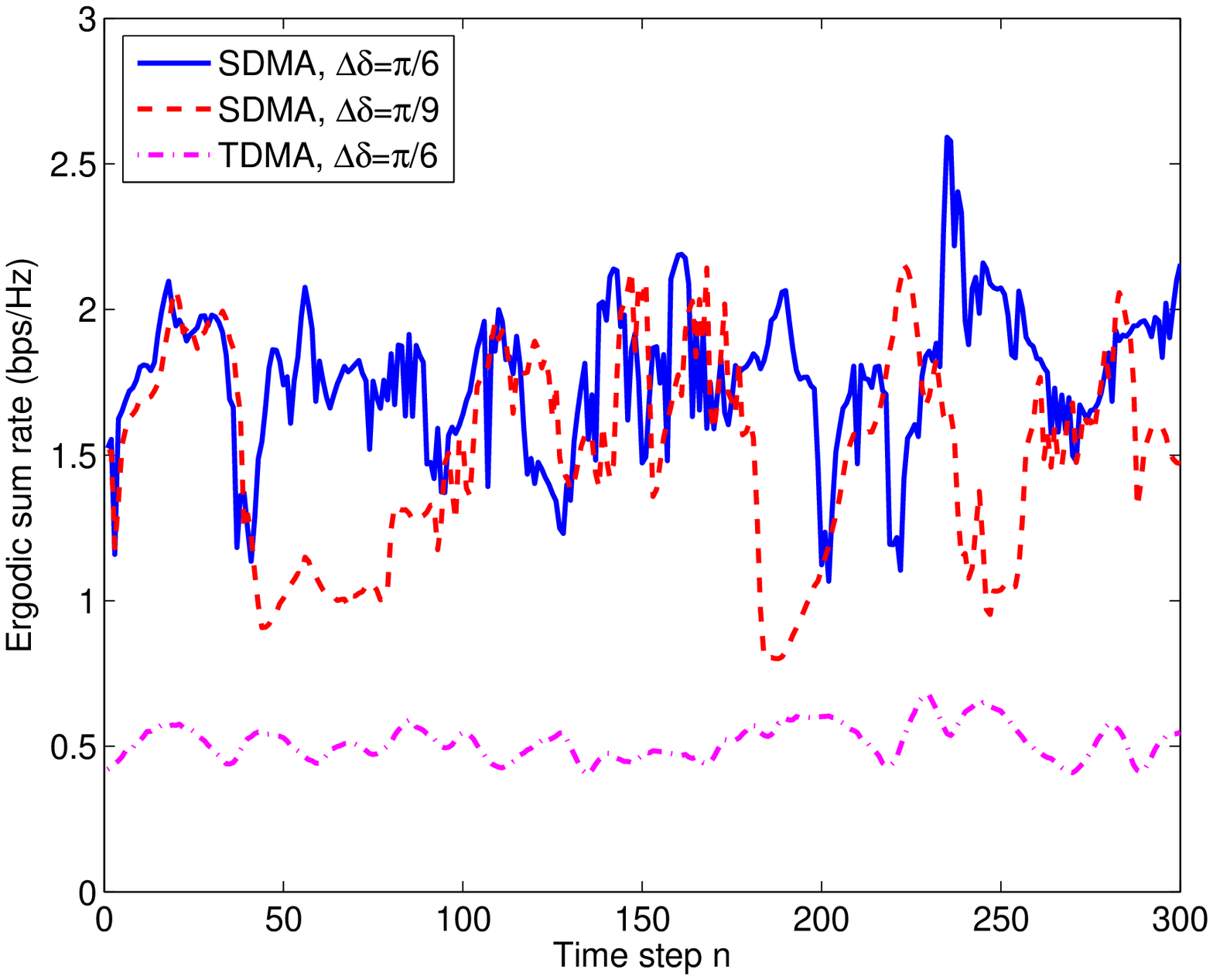}
\caption{Comparison of sum rate performance (bps/Hz) with $K=10$ and $\frac{P_t}{\sigma^2}=45\textrm{dB}$, proportional fair. The average sum rates are: $1.6968\;(\textrm{SDMA}, \Delta\delta=\frac{\pi}{6})$, $1.6042\;(\textrm{SDMA}, \Delta\delta=\frac{\pi}{9})$, $0.5139\;(\textrm{TDMA}, \Delta\delta=\frac{\pi}{6})$.}\label{f8}
\end{figure}

Figs.~\ref{f7}-\ref{f8} show the ergodic sum rate for the different
scenarios. For each time step, the rate is calculated by averaging
over 1000 independent channel realizations.  Results for both
$\Delta\delta=\frac{\pi}{6}$ and $\frac{\pi}{9}$ are plotted.
Increasing the maximum turning rate will clearly provide better
performance since it decreases the extra distance that must be flown
to complete a loop maneuver and the amount of time that the 
array is aligned with the angle-of-arrival of each user's
signal (where the ability of the array to suppress interference is
minimized). The benefit of using SDMA is also apparent from
Figs.~\ref{f7}-\ref{f8}, where we see that a rate gain of
approximately a factor of 3.3 is achieved over the TDMA scheme. We
also note that the obtained sum rate is only about $15\%$ less than
what would be achieved assuming no interference, indicating the
effectiveness of the beamforming algorithm.

\begin{figure}
\centering
\includegraphics[height=3.5in, width=4.8in]{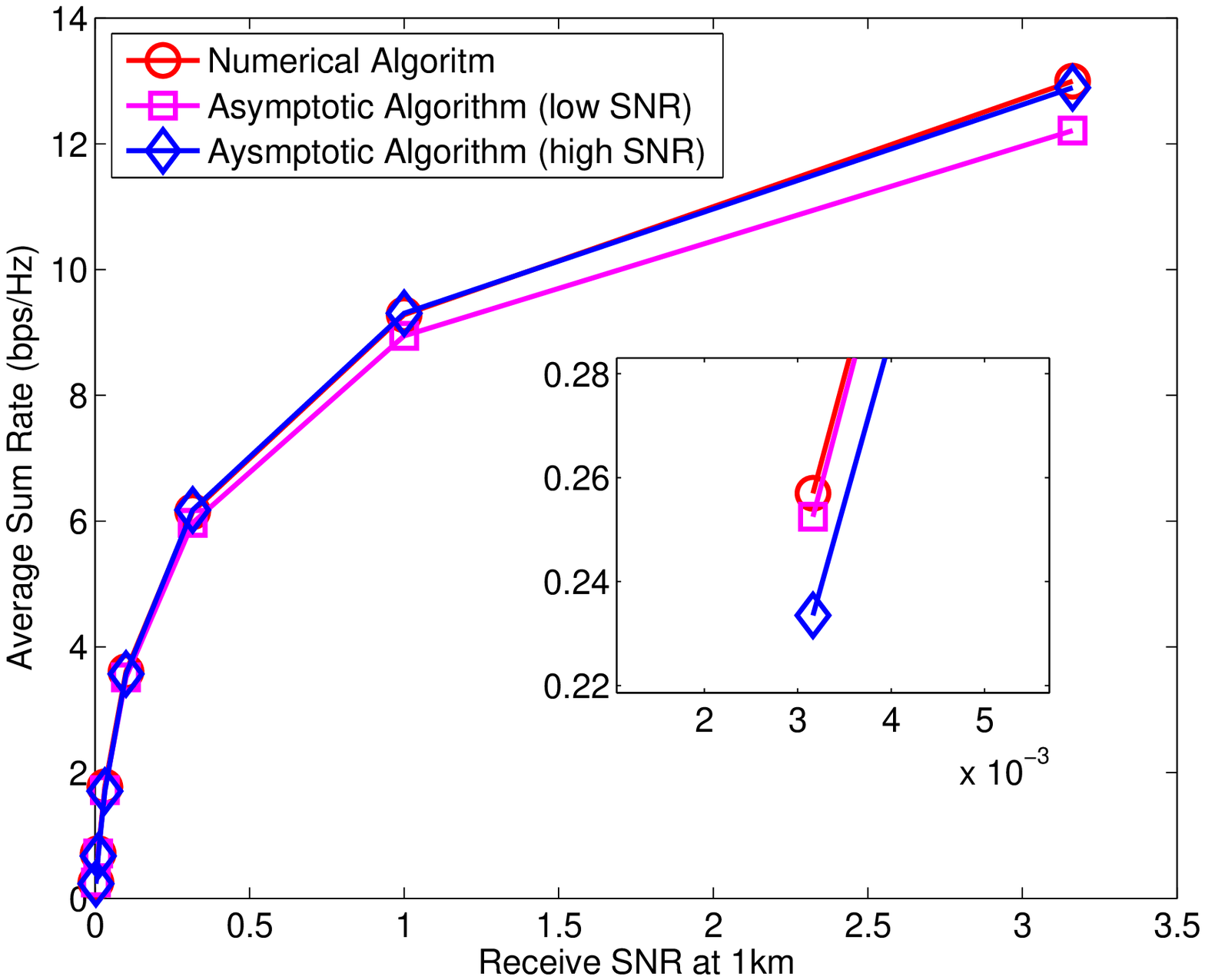}
\caption{Comparison of the average sum rate of 
the line-search and closed-form approximations with
$\Delta\delta=\frac{\pi}{9}$, $K=1000$, maximizing sum rate. The x-axis denotes the SNR 
that would be observed at the UAV for a user node at a distance of $1\textrm{km}$.} 
\label{f12}
\end{figure}

Fig.~\ref{f12} compares the average sum rate of the line-search
algorithm in~(\ref{eq:opt2}) with both the low- and high-SNR
approximations derived in the previous section. The performance is
plotted as a function of the received SNR that would be observed at
the UAV from a ground node located at a distance of $1\textrm{km}$. Although the
approximate algorithms were derived separately under different
SNR assumptions, both of them yield performance essentially
identical to~(\ref{eq:opt2}) over all SNR values.  Each approximate
algorithm is slightly better than the other in its respective
SNR regime, but the performance difference is small.

\section{Conclusion}\label{sec:conclude}

We have investigated the problem of positioning a multiple-antenna UAV
for enhanced uplink communications from multiple ground-based users.
We studied the optimal UAV trajectory for a case involving two static
users, and derived an approximate method for finding this trajectory
that only requires a simple line search.  For the case of a network of
mobile ground users, an adaptive heading algorithm was proposed that
uses predictions of the user terminal positions and beamforming at the
UAV to maximize SINR at each time step.  Two kinds of optimization
problems were considered, one that maximizes a lower bound on the
average uplink sum rate and one that guarantees fairness among the
users using the proportional fair method. Simulation results indicate
the effectiveness of the algorithms in automatically generating a
suitable UAV heading for the uplink network, and demonstrate the
benefit of using SDMA over TDMA in achieving the best throughput
performance.  We also derived approximate solutions to the UAV heading
problem for low- and high-SNR scenarios; the approximations allow for
a closed-form solution instead of a line search, but still provide
near-optimal performance in their respective domains.



\appendices

\section{Derivation of UAV Trajectory for Two-User Case}

In this appendix, we find an approximation to the problem posed
in equation~(\ref{eq:trajecopt}):
\begin{eqnarray}
\max_{\delta,C_a,C_b}&&\bar{R}\\
\hspace{-1.5em}\mathrm{subject\quad to}&& 0 \le \delta \le \frac{\pi}{2}\nonumber\\
\hspace{-1.5em}&& C_{\min}\le C_b \le C_a \le C_{\max},\nonumber
\end{eqnarray}
where $\bar{R}$ is defined in~(\ref{eq:Rbar}).  To begin with, we
observe that, due to the symmetric trajectory centered at the midpoint
between the two ground nodes, the expected data rate averaged over the
trajectory will be the same for both users:
\[ 
\int_{\mathcal{C}} \log_{2}(1+SINR_1(p)) dp = 
\int_{\mathcal{C}} \log_{2}(1+SINR_2(p)) dp \; .
\]
Thus, we can focus on evaluating the SINR for just one of the users.
For large $K$, we can ignore the Rayleigh component of the channel,
and assume that $\mathbf{h}_i^{'} \approx \bar{\mathbf{h}_i}$.  We
replace the explicit dependence of the channel on $n$ with an implicit
dependence on a point $p$ along the trajectory defined in
Fig.~\ref{trajectory}.  At point $p$, the SINR for user~1 can be
expressed as
\begin{eqnarray}\label{eq:SINR1}
SINR_{1}&=&\frac{P_t}{d_1^{\alpha}}\bar{\mathbf{h}}_1^{H}\left(\sigma^2\mathbf{I}_{M}+\frac{P_t}{d_2^{\alpha}}\bar{\mathbf{h}}_2\bar{\mathbf{h}}_2^H\right)^{-1}\bar{\mathbf{h}}_1\nonumber\\
&=&\frac{MP_t}{d_1^{\alpha}\sigma^2}-\frac{P_t^2}{d_1^{\alpha}d_2^{\alpha}\sigma^4}\frac{\left|\bar{\mathbf{h}}_1^H\bar{\mathbf{h}}_2\right|^2}{1+\frac{MP_t}{d_2^{\alpha}\sigma^2}},
\end{eqnarray}
where 
\begin{eqnarray}\label{eq:correlation}
\left|\bar{\mathbf{h}}_1^H\bar{\mathbf{h}}_2\right|&=&\left|\frac{\sin\Big(\frac{M\pi}{2}\big(\cos(\phi_1)\sin(\theta_{1})-\cos(\phi_2)\sin(\theta_{2})\big)\Big)}{\sin\Big(\frac{\pi}{2}\big(\cos(\phi_1)\sin(\theta_{1})-\cos(\phi_2)\sin(\theta_{2})\big)\Big)}\right|,
\end{eqnarray}
and $\cos(\phi_i)$ and $\sin(\theta_i)$ are defined in~(\ref{eq:aoa}).
Note that in addition to $\bar{\mathbf{h}}_1$, the parameters $d_i,
\phi_i$ and $\theta_i$ all implicitly depend on $p$.  

Using Jensen's inequality, the following upper bound for $\bar{R}$ 
can be found:
\begin{eqnarray}\label{eq:rateub}
\bar{R}&\le\log_{2}(1+\mathbb{E}\{SINR_1\})+\log_{2}(1+\mathbb{E}\{SINR_2)\}.
\end{eqnarray}
We will proceed assuming that an operating point that maximizes the
upper bound will also approximately optimize $\bar{R}$.
Based on~(\ref{eq:SINR1}) and assuming we have a high SNR scenario
where $\frac{P_t}{d_i^{\alpha}\sigma^2}\gg 1$, 
\begin{eqnarray}\label{eq:sinrapprox}
\mathbb{E}\{SINR_1\}&\overset{(d)}{\approx}&\frac{P_t}{\sigma^2}\mathbb{E}
\left\{\frac{M}{d_1^{\alpha}}-\frac{|\mathbf{h}_1^H\mathbf{h}_{2}|^2}{d_1^{\alpha}M}\right\}\nonumber\\
&\overset{(e)}{\approx}&\frac{P_t}{d_1^{\alpha}\sigma^2}
\left(M-\frac{\mathbb{E}\{|\mathbf{h}_{1}^H\mathbf{h}_{2}|^2\}}{M}\right),
\end{eqnarray}
where (d) is due to the high SNR assumption and (e) follows from the
assumption that $C_{\max}\ll d_1$.  The dependence of $SINR_1$ on $d_2$
is thus eliminated, and in what follows we drop the subscript on $d_1$
and write it simply as $d$.  

Substituting equation~(\ref{eq:sinrapprox}) in~ (\ref{eq:rateub}), and
replacing the objective function in problem~(\ref{eq:trajecopt}) with
the upper bound of~(\ref{eq:rateub}), our optimization problem is
approximately given by
\begin{eqnarray}\label{eq:trajecopt2}
\max_{\delta,C_a,C_b}&&\log_2\left(1+\frac{MP_t}{d^{\alpha}\sigma^2}-\frac{P_t\mathbb{E}\{|\mathbf{h}_{1}^H\mathbf{h}_{2}|^2\}}{Md^{\alpha}\sigma^2}\right)\\
\hspace{-1.5em}\mathrm{subject\quad to}&& 0 \le \delta \le \frac{\pi}{2}\nonumber\\
\hspace{-1.5em}&& C_{\min} \le C_b \le C_a \le C_{\max}.\nonumber
\end{eqnarray}
Since the objective function in (\ref{eq:trajecopt2}) is monotonically
decreasing with $\mathbb{E}\{|\mathbf{h}_{1}^H\mathbf{h}_{2}|^2\}$, an
equivalent problem is formulated as
\begin{eqnarray}\label{eq:trajecopt3}
\min_{\delta,C_a,C_b}&&\mathbb{E}\{|\mathbf{h}_{1}^H\mathbf{h}_{2}|^2\}\\
\hspace{-1.5em}\mathrm{subject\quad to}&& 0 \le \delta \le \frac{\pi}{2}\nonumber\\
\hspace{-1.5em}&& C_{\min} \le C_b \le C_a \le C_{\max}.\nonumber
\end{eqnarray}
The interpretation of~(\ref{eq:trajecopt3}) is that the optimal
trajectory minimizes the average correlation between the two users' 
channels.

The calculation of $\mathbb{E}\{|\mathbf{h}_{1}^H\mathbf{h}_{2}|^2\}$
includes the integral of the function
\[ \frac{\sin^2\Big(\frac{M\pi}{2}\big(\cos(\phi_1)\sin(\theta_{1})-\cos(\phi_2)\sin(\theta_{2})\big)\Big)}
{\sin^2\Big(\frac{\pi}{2}\big(\cos(\phi_1)\sin(\theta_{1})-\cos(\phi_2)\sin(\theta_{2})\big)\Big)}
\]
with respect to $p$, which is difficult to evaluate. To
simplify~(\ref{eq:trajecopt}), we assume that, compared with the distance to the
users on the ground, the UAV moves over a small region, and for
purposes of analyzing the mathematics, one can assume that the
UAV essentially remains fixed at the midpoint between the
two users.  Only the heading of the UAV changes the uplink
rate in this case. Under this assumption, the elevation angles $\phi_1$, $\phi_2$ are constant and equal $\phi_1=\phi_2=\phi^{'}$, and the azimuth angles $\theta_1$, $\theta_2$ are piecewise constant. When UAV flies along
$C_a$, they are equal to $\theta_1$ and $\theta_2$; when the UAV flies along $C_b$, they are equal to $\theta_1+\frac{\pi}{2}$, $\theta_2+\frac{\pi}{2}$. Note that since $\theta_2=\theta_1+\pi$ always holds, then $\sin(\theta_2)=-\sin(\theta_1)$ and we have

\begin{equation}\label{eq:ca}
|\mathbf{h}_1^H\mathbf{h}_2|^2=\frac{\sin^2(M\pi\cos(\phi^{'})\sin(\theta_1))}{\sin^2(\pi\cos(\phi^{'})\sin(\theta_1))}.
\end{equation}
Note also that $\theta_1+\delta=\frac{\pi}{2}$, and hence $\sin(\theta_1)=\cos(\delta)$. Thus
\begin{equation}\label{eq:ca}
|\mathbf{h}_1^H\mathbf{h}_2|^2=\frac{\sin^2(M\pi\cos(\phi^{'})\cos(\delta))}{\sin^2(\pi\cos(\phi^{'})\cos(\delta))}.
\end{equation}
Along $C_a$, the UAV flies with heading $\delta$ and along $C_b$, the
UAV flies with heading $\delta+\frac{\pi}{2}$, so that
$\cos(\delta+\frac{\pi}{2})=-\sin(\delta)$. Thus, we have
\begin{eqnarray}\label{eq:corre}
\mathbb{E}\{|\mathbf{h}_1^H\mathbf{h}_2|^2\}=\frac{C_a}{C_a+C_b}\frac{\sin^2(M\pi\cos(\phi_{i}^{'})\cos(\delta))}{\sin^2(\pi\cos(\phi_{i}^{'})\cos(\delta))}+
\frac{C_b}{C_a+C_b}\frac{\sin^{2}(M\pi\cos(\phi_{i}^{'})\sin(\delta))}{\sin^{2}(\pi\cos(\phi_{i}^{'})\sin(\delta))}.
\end{eqnarray}
Substituting~(\ref{eq:corre}) into the objective function
of problem~(\ref{eq:trajecopt3}) yields
\begin{eqnarray}\label{eq:trajecopt4}
\min_{\delta,C_a,C_b}&&\frac{C_a}{C_a+C_b}\frac{\sin^2(M\pi\cos(\phi_{i}^{'})\cos(\delta))}{\sin^2(\pi\cos(\phi_{i}^{'})\cos(\delta))}+
\frac{C_b}{C_a+C_b}\frac{\sin^{2}(M\pi\cos(\phi_{i}^{'})\sin(\delta))}{\sin^{2}(\pi\cos(\phi_{i}^{'})\sin(\delta))}\\
\hspace{-1.5em}\mathrm{subject\quad to}&& 0 \le \delta \le \frac{\pi}{2}\nonumber\\
\hspace{-1.5em}&& C_{\min} \le C_b \le C_a \le C_{\max}.\nonumber
\end{eqnarray}

We now show that Problem (\ref{eq:trajecopt4}) is equivalent to an
optimization problem over the single variable $\delta$. 
First define
\begin{eqnarray}
s_1 & = & \frac{\sin^{2}(M\pi\cos(\phi_{i}^{'})\cos(\delta))}{\sin^{2}(\pi\cos(\phi_{i}^{'})
\cos(\delta))}\\
s_2 & = & \frac{\sin^{2}(M\pi\cos(\phi_{i}^{'})\sin(\delta))}{\sin^{2}(\pi\cos(\phi_{i}^{'})\sin(\delta))} \\
R_c & = & \frac{C_{\max}}{C_{\min}}\\
R & = & \frac{C_a}{C_b} \; ,
\end{eqnarray}
so that $1\le R \le R_c$.  Then the objective function of
(\ref{eq:trajecopt4}) can be rewritten as
\begin{equation}
\frac{R}{1+R}s_1+\frac{1}{1+R}s_2=s_1+\frac{s_2-s_1}{1+R}.
\end{equation}
Given a heading direction $\delta\in[0,\frac{\pi}{2}]$, if $s_2\ge
s_1$, then the objective function is minimized when $R=R_c$.  Otherwise, if $s_2<s_1$,
$R=1$ minimizes the objective function. The domain
$[0,\frac{\pi}{2}]$ can be divided into two sets $\mathcal{S}_1$ and
$\mathcal{S}_2$, such that for $\delta\in\mathcal{S}_1$, $s_2<s_1$,
and for $\delta\in\mathcal{S}_2$, $s_2\ge s_1$. Then problem
(\ref{eq:trajecopt4}) can be divided into two subproblems
\begin{eqnarray}\label{eq:optsimple2}
\min_{\delta}&&\frac{R_c}{1+R_c}s_1+\frac{1}{1+R_c}s_2\\
\hspace{-1.5em}\mathrm{subject\quad to}&& \delta\in\mathcal{S}_2.\nonumber
\end{eqnarray}
\begin{eqnarray}\label{eq:optsimple3}
\min_{\delta}&&\frac{1}{2}s_2+\frac{1}{2}s_1\\
\hspace{-1.5em}\mathrm{subject\quad to}&& \delta\in\mathcal{S}_1.\nonumber
\end{eqnarray}
Since $s_1(\frac{\pi}{2}-\delta)=s_2(\delta)$, for each $\delta\in\mathcal{S}_2$, we have $\frac{\pi}{2}-\delta\in\mathcal{S}_1$ and vice versa. Thus the following equation holds
\begin{equation}
\frac{R_c}{1+R_c}s_1(\delta)+\frac{1}{1+R_c}s_2(\delta)<\frac{1}{2}s_1(\delta)+\frac{1}{2}s_2(\delta)=\frac{1}{2}s_2\left(\frac{\pi}{2}-\delta\right)+\frac{1}{2}s_1\left(\frac{\pi}{2}-\delta\right).
\end{equation}
Then the minimum value of (\ref{eq:optsimple2}) must be smaller than the minimum value of (\ref{eq:optsimple3}) and problem (\ref{eq:trajecopt4}) is equivalent to problem (\ref{eq:optsimple2}). For each $\delta\in\mathcal{S}_2$, the following equation holds
\begin{equation}
\frac{R_c}{1\!+\!R_c}s_1(\delta)\!+\!\frac{1}{1\!+\!R_c}s_2(\delta)\!<\!\frac{R_c}{1\!+\!R_c}s_2(\delta)\!+\!\frac{1}{1\!+\!R_c}s_1(\delta)\!=\!\frac{R_c}{1\!+\!R_c}s_1\!\left(\frac{\pi}{2}\!-\!\delta\right)\!+\!\frac{1}{1\!+\!R_c}s_2\!\left(\frac{\pi}{2}\!-\!\delta\right),
\end{equation}
and problem~(\ref{eq:optsimple2}) is thus equivalent to  
\begin{eqnarray}\label{eq:trajecopt5}
\min_{\delta}&&\frac{R_c}{1+R_c}s_1+\frac{1}{1+R_c}s_2\\
\hspace{-1.5em}\mathrm{subject\quad to}&& 0<\delta<\frac{\pi}{2} \; . \nonumber
\end{eqnarray}
Equation (\ref{eq:optsimple}) follows directly from (\ref{eq:trajecopt5}).


%

\bibliography{reference}

\newpage
\input{twouser.TpX}

\end{document}